\documentclass[aps,prc,nofootinbib,reprint,preprintnumbers,superscriptaddress]{revtex4-1}
\usepackage{supertabular}
\usepackage{lipsum}
\usepackage{blindtext}
\usepackage{amsfonts}
\usepackage{tensor}
\usepackage{graphicx}
\usepackage{pict2e}
\usepackage{balance}
\usepackage{blindtext}

\usepackage{amssymb, amsmath,environ}

\usepackage{color}
\usepackage[colorlinks=true,urlcolor=black,linkcolor=black,citecolor=red]{hyperref}


\newcommand{\bse}{\begin{subequations}}
	\newcommand{\ese}{\end{subequations}}
\newcommand{\be}{\begin{equation}}
\newcommand{\ee}{\end{equation}}

\NewEnviron{eqsplit}{%
	\begin{equation}\begin{split}
	\BODY
	\end{split}\end{equation}
}

\makeatletter
\newcommand*\bigcdot{\mathpalette\bigcdot@{.5}}
\newcommand*\bigcdot@[2]{\mathbin{\vcenter{\hbox{\scalebox{#2}{$\m@th#1\bullet$}}}}}
\makeatother

\newcommand{\bea}{\begin{eqnarray}}
\newcommand{\eea}{\end{eqnarray}}
\newcommand{\ba}{\begin{array}}
	\newcommand{\ea}{\end{array}}

\begin{document}
%\preprint{MITP-23-014}		

\title{Random model of flow decorrelation}

\author{Piotr Bo\.zek}
\email[]{Piotr.Bozek@fis.agh.edu.pl}
\affiliation{AGH University of Krakow, Faculty of Physics and Applied Computer Science, al. Mickiewicza 30, 30-059 Cracow, Poland}

\author{Hadi Mehrabpour}
\email[]{mehrabph@uni-mainz.de}
\affiliation{PRISMA$^{+}$ Cluster of Excellence \& Mainz Institute for Theoretical Physics,
	Johannes Gutenberg-Universit\"at Mainz, 55099 Mainz, Germany}

\begin{abstract}
  The collective flow generated in relativistic heavy-ion collisions fluctuates from event to event.  The fluctuations lead to a decorrelation of  flow vectors measured in separate bins in phase space. These effects have been measured in experiments and observed in numerical simulations in hydrodynamic models.
  We present a simple random model of flow decorrelation in pseudorapidity. Analytical expressions for the flow factorization breaking coefficients for flow vectors, flow vector magnitudes, and flow  angles are derived. The model explains the relations between different factorization breaking coefficients found in experimental data and model simulations. In particular, it  is found  that  the flow angle decorrelation constitutes about one half of the total flow vector decorrelation.
\end{abstract}


\maketitle
\section{Introduction}

One of the methods of  investigation of the hot and dense matter created in the
interaction region of relativistic heavy-ion collisions is the analysis of the collective flow from the spectra of emitted particles  \cite{Huovinen:2006jp,Voloshin:2008dg,Hirano:2008hy,Heinz:2009xj,Heinz:2013th}. Strong pressure gradients in the  fireball cause a rapid expansion. The azimuthal
asymmetry of the collective flow can be quantified using the harmonic flow
coefficients, describing the magnitude and the azimuthal direction of the flow.
The dynamical model commonly used to describe the generation of the collective flow in the
rapid expansion of the source is the relativistic viscous hydrodynamics
\cite{Schenke:2010rr}.

The collective flow generated   reflects the properties
of the initial state of the collision. The initial conditions fluctuate
from event to event and the final flow observables fluctuate as well  \cite{Aguiar:2001ac,Takahashi:2009na,Alver:2010gr,Schenke:2010rr,Schenke:2012wb}. One of the aspects of such fluctuations is the decorrelation of the flow harmonics measured in different pseudorapidity bins. Model calculations
predict  a deviation from unity of the correlation coefficient between two
flow vectors in
 separate pseudorapidity or transverse momentum  bins \cite{Bozek:2010vz,Gardim:2012im}. This correlation coefficient of flow vectors is called the flow factorization breaking coefficient. It has been measured in experiments \cite{CMS:2015xmx,ATLAS:2017rij,ATLAS:2020sgl,Nie:2019bgd,ALICE:2022dtx} and qualitatively reproduced  in models \cite{Gardim:2012im,Kozlov:2014fqa,Gardim:2017ruc,Zhao:2017yhj,Bozek:2018nne,Barbosa:2021ccw,Bozek:2010vz,Jia:2014ysa,Pang:2015zrq,Sakai:2021rug,Cimerman:2021gwf}. In this paper we study the decorrelation of the collective flow for different bins in pseudorapidity.

Correlations of higher moments of flow vectors in different phase-space bins have been studied as well, both in experiment \cite{ATLAS:2017rij,ALICE:2022dtx} and in models \cite{Jia:2014ysa,Bozek:2017qir,Wu:2018cpc,Bozek:2018nne,Bozek:2021mov}.
The measurement of the decorrelation using four-particle correlators allows
one to estimate separately the flow vector magnitude and flow angle
decorrelation in different bins. Experimental and model calculation show that the total flow vector decorrelation is composed in approximately equal parts
from the flow magnitude and flow angle decorrelation.
The second observation is that the decorrelation of higher powers of
flow vectors is stronger than the decorrelation of simple flow vectors. The  factorization breaking coefficient for the second or third power of the flow vector can be approximated  as the second or third power of the simple flow vector factorization breaking coefficient.
Finally, it has been observed in model calculations that the decorrelation
is the strongest for events (or classes of events) where the overall flow
is the smallest.

These effects have been observed  in numerical simulations,
but no simple understanding has been given. In this paper, we present a simple model of flow decorrelation in different phase-space bins. We assume that the
flow vector in a small pseudorapidity bin can be written as
a sum of the overall flow
(averaged over the whole event) and of a random vector component. Assuming the
independence of the directions  of random component  of the flow and
of the average flow, the model can explain qualitatively the effects
observed in model simulations and in the experimental data.
We present a number of analytical results for the factorization coefficients
for flow vectors, for  powers of flow vectors, for  flow magnitudes
and for  flow angles. We show  that  qualitatively similar relations between different factorization breaking coefficients are found  in simulations using a realistic hydrodynamic model.

We study three- and four-bin measures of the flow decorrelation in pseudorapidity, used
in experimental analyses \cite{CMS:2015xmx,ATLAS:2017rij}.
The random model
of flow decorrelation can describe qualitatively the relations between
different three- and four-bin measures of  the flow vector, flow magnitude, and flow angle
decorrelation observed in model calculations and in the experimental data. Also in this case a relation between the decorrelation of the second or  third power  of flow vectors and the second or third power of the simple flow vector decorrelation is found.

The flow factorization breaking effect is briefly described in Sec. \ref{sec3}.
Our model of the flow decorrelation due to a random component in the flow vector is introduced in Sec. \ref{sec:random}. In the following sections applications of the random flow decorrelation model for the   flow decorrelation in pseudorapidity are presented: the factorization breaking coefficient of the flow in two bins in pseudorapidity (Sec. \ref{sec:facbreak}), the analysis of flow angle decorrelation (Sec. \ref{sec:flowangle}), and the calculations for three- and four-bin measures of flow deccorelation (Sec. \ref{sec:34bins}).  The results are summarized
in the last section.

\section{Flow correlation}\label{sec3}

 \begin{figure*}[t!]
 	\begin{center}
 		\begin{tabular}{cc}
 			\includegraphics[scale=0.48]{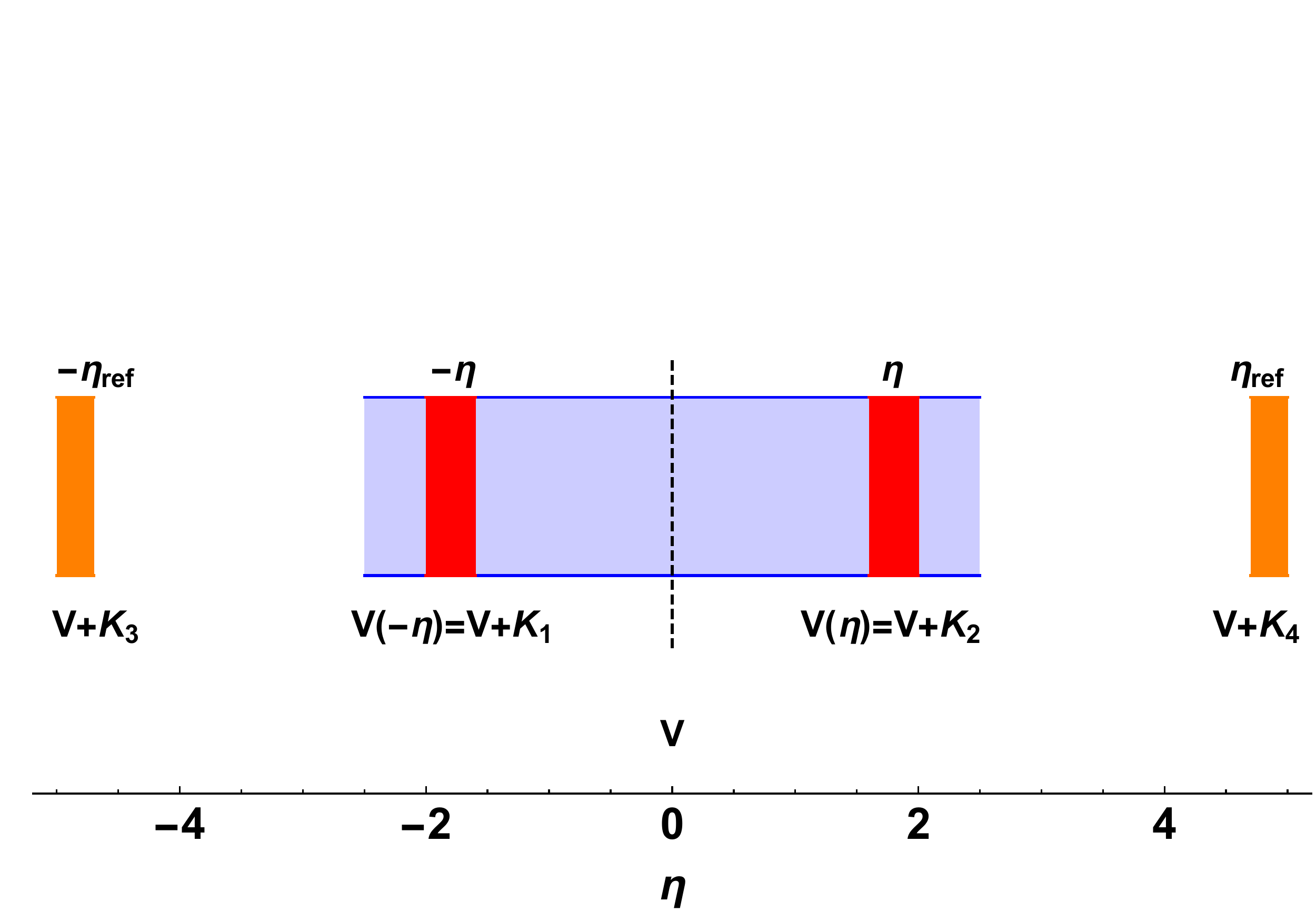}			
 		\end{tabular}		
 		\caption{Pseudorapidity bins used in  definitions of  factorization breaking coefficients of the collective flow (two, three, or all four bins depending on the case).} 
 		\label{fig:bins}
 	\end{center}
 \end{figure*}

The flow in an event can be defined using the harmonic coefficients of the azimuthal distribution of  emitted particles. We use the notation $V_n= v_n e^{in\Psi_n}$, where $v_n$ and $\Psi_n$ are the magnitude and event-plane angle for the $n$th order harmonic flow. The flow vector $V_n$ cannot be reconstructed in each event. Rotationally invariant combinations of moments of flow vectors can be estimated from the moments of the corresponding $q_n$ vectors in a phase-space region $A$,
\begin{equation}\label{eq:qvector}
q_n(A)=\frac{1}{N}\sum_{k \in A} e^{in\phi_k},
\end{equation}
where the sum runs over all $N$ hadrons in the selected phase-space region $A$, and $\phi_k$ are their azimuthal angles.
The event average of the $q_n$ vector moments is an estimator of the corresponding moments of the flow vectors, e.g.,
\begin{equation}\label{eq:moments}
\langle (V_n)^{m}(V_n^*)^{m}\rangle = \langle q_n^m (q_n^\star)^m \rangle,
\end{equation}
where the angular brackets denote an average  over the events. 

The factorization breaking of the collective flow means that the flow moment calculated in different regions in phase space ($A$ and $B$),
\begin{equation}
\langle V_n(A)V_n^*(B)\rangle = \frac{1}{N_A N_B}\sum_{k \in A, j\in B } e^{i n(\phi_k-\phi_j)},
\end{equation}
does not factorize into flow moments calculated  \cite{Bozek:2010vz,Gardim:2012im} in the same bin,
\begin{equation}
\langle V_n(A)V_n^*(A)\rangle =  \frac{1}{N_A (N_A-1)}\sum_{k\neq j  \in A } e^{i n(\phi_k-\phi_j)} \ ,
\end{equation}
which gives
\begin{equation}
\langle V_n(A)V_n^*(B)\rangle \neq \sqrt{ \langle V_n(A)V_n^*(A)\rangle  \langle V_n(B)V_n^*(B)\rangle} \ .
\end{equation}
The factorization breaking coefficient is the correlation coefficient of flow vectors in two phase-space regions
\begin{equation}
\rho_{V_n}(A,B)=\frac{\langle V_n(A)V_n^*(B)\rangle}{\sqrt{ \langle V_n(A)V_n^*(A)\rangle \langle V_n(B)V_n^*(B)\rangle}} .
\label{eq:facb}
\end{equation}
If the flow dominates the multiparticle correlation we have $\rho_{V_n}(A, B)\leq1$.
The flow correlation coefficient (factorization breaking coefficient) can be used as a measure of flow decorrelation for two bins in pseudorapidity \cite{Bozek:2010vz} or in transverse momentum \cite{Gardim:2012im}.
The factorization breaking coefficient $r_n(p_1,p_2)$ in transverse momentum can be measured in experiment \cite{CMS:2015xmx,CMS:2013bza,Zhou:2014bba} and calculated in models \cite{Gardim:2012im,Kozlov:2014fqa,Gardim:2017ruc,Zhao:2017yhj,Bozek:2018nne,Barbosa:2021ccw}.

The factorization breaking coefficient  in pseudorapidity, Eq. (\ref{eq:facb}), contains a significant contribution from nonflow effects \cite{Bozek:2010vz}. 
A modified factorization breaking coefficient has been proposed \cite{CMS:2015xmx}, defined as a ratio of two flow vector covariances taken for two different pairs of bins (Fig. \ref{fig:bins}),
\begin{equation}
\mathcal{R}^{(1)}_{n;V}=\frac{\langle V_n(-\eta)V_n^*(\eta_{ref})\rangle}{\langle V_n(\eta)V_n^*(\eta_{ref})\rangle},
\label{eq:r3bin}
\end{equation}
where $\eta_{ref}$ is the reference pseudorapidity common to the numerator and the denominator. The bins are placed such that both $|\eta_{ref}-\eta|$ and $|\eta_{ref}+\eta|$ are large enough to suppress nonflow correlations. 
The experimentally measured \cite{CMS:2015xmx,Huo:2017hjv,ATLAS:2017rij} decorrelation in pseudorapidity using Eq.~(\ref{eq:r3bin}) can be qualitatively reproduced in hydrodynamic and cascade models \cite{Bozek:2015bha,Pang:2015zrq,Xiao:2015dma}.
Besides the simple factorization breaking coefficient defined in  Eq.~(\ref{eq:facb}), factorization breaking coefficients for higher moments of flow vectors can be defined \cite{Jia:2017kdq,ATLAS:2017rij,Bozek:2018nne,Bozek:2021mov,ALICE:2022dtx}.
Based on these factorization-breaking coefficients of higher order flow moments, the flow magnitude and flow angle decorrelation can be estimated separately \cite{Jia:2014ysa,Bozek:2018nne}.
In the following, we discuss a simple model of flow decorrelation that explains qualitatively the relations between different factorization breaking coefficients. To illustrate the effects in model calculations, we use a 3+1-dimensional viscous hydrodynamic model \cite{Schenke:2010rr} with particle emission through statistical hadronization \cite{Chojnacki:2011hb} at freeze-out. Details of the three-dimensional fluctuating initial conditions and hydrodynamic model can be found in Ref \cite{Bozek:2017qir}.
In the present work, we use  Pb+Pb collisions at $\sqrt{s_{NN}}=5.02$TeV for two centralities, $0-5$\% and $30-40$\%, as a numerical example of the analytical identities discusses in the random model of flow decorrelation.

\section{Random model of flow decorrelation}

\label{sec:random}

The flow  in the same event, but in two separate phase-space regions, is slightly decorrelated. To understand the basic features of this effect, we study a simple random flow decorrelation model. The model is simple enough so that, in the limit of small flow decorrelations, many observables, such as the factorization breaking coefficients and flow correlations, can be estimated analytically.

In a given event, the flow measured in two specific bins differs. Also, the flow $V_i=V+K_i$ defined in a small bin $i$ in phase space differs from the flow $V$ averaged over the whole acceptance region  depicted in Fig.~\ref{fig:bins}. The flow in a small bin  is composed of the average flow in the whole acceptance region and a random component $K_i$. Both $V$ and $K_i$ have nontrivial event-by-event distributions. In the following, we assume  that all the small bins considered are of the same size and that in each bin $i$ the flow distribution  is the same. It is a good approximation for the flow defined in pseudorapidity bins. Alternatively,  one can transfer the definition of the model to the scaled flow $V_i/\langle V_i\rangle$, e.g., for the modeling of flow factorization breaking in transverse momentum. 

The combined probability distribution of two components of the flow $V$ and $K_i$, $i=1,\dots, N$ can be very complicated in principle. We make one simplifying assumption:
\begin{equation}
\langle V^m (K_i^\star)^m \rangle =0 \ \ , \ \textrm{}\ \ m=1,2,\dots \ .
\end{equation}
We have checked quantitatively the magnitude of the above term in comparison to the terms that are kept in the expansion.  We find
          that the approximation works well for the triangular flow and for the elliptic flow in central collisions ($|\langle V K^\star \rangle | < 0.05 \langle K K^\star \rangle$ ). The approximation is not as good, but still acceptable, for the elliptic flow in semicentral collisions  ($|\langle V K^\star \rangle | \simeq 0.13 \langle K K^\star \rangle$ ).
Note that the above properties are fulfilled under the assumption of a random relative orientation of flow vectors $V$ and $K_i$. On the other hand, the magnitudes could still be correlated,
\begin{equation}
\langle v^m |K_i|^l \rangle \neq 0 \ .
\end{equation}
The local random component of the flow $K_i$ fulfills the constraint
\begin{equation}
\langle (V+K_i)(V+K_i)^\star \rangle = Var(V)+ Var(K_i)= Var(V) + C^2 \ ,
\label{eq:varvk}
\end{equation}
where $C^2$ is a constant defined as the difference between flow vector variance in a small bin and flow vector variance in the whole acceptance region. Nontrivial flow correlation between two different bins is encoded in the correlation of the random components $K_i$ and $K_j$ of the flow in the two bins. Phenomenologically, one expects that as the separation between the two bins increases the random components of flow vectors $K_i$ and $K_j$ become less aligned. This means that the covariance
\begin{equation}
\langle K_i K_j^\star \rangle
\end{equation}
decreases with increasing bin separation. Another correlation between random components comes from the global constraint
\begin{equation}
\sum_i K_i = 0 \ . \label{eq:glsum}
\end{equation}
For the study of the  covariances between flows in two bins, we define
\begin{equation}
A=\frac{K_i+K_j}{2}\quad \text{and} \quad\Delta=\frac{K_i-K_j}{2} \ .
\end{equation}
Note that, due to the properties of $K_i$ and $K_j$, both $A$ and $\Delta$ depend on the separation between the bins. The distribution of $\Delta$ gets wider as the bin separation increases (reflecting the increasing decorrelation of $K_i$ and $K_j$). On the other hand, the distribution of $A$ gets narrower,
\begin{equation}
\langle AA^\star\rangle= C^2 -\langle \Delta\Delta^\star \rangle,
\end{equation}
due to the constraint from Eq.~(\ref{eq:varvk}). We have 
\begin{equation}
\langle A \Delta\rangle=0 \ .
\end{equation}
For bins in pseudorapidity, we specify $V(\eta)=V+K_i$ for the flow vector in a bin of fixed size around the pseudorapidity $\eta$. As indicated above, the distribution of $A$ and $\Delta$ depends only on the pseudorapidity separation $|\eta_1-\eta_2|$ between the two bins.
Besides the mathematical constraint in Eq.~(\ref{eq:glsum}) or global momentum conservation constraints \cite{Dasgupta:2022psm}, the correlations between the two random components $K_i$ and $K_j$ can have a nontrivial physical origin. In the hydrodynamic model, fluctuations in the initial distribution in space-time rapidity generate fluctuations in the final flow in pseudorapidity \cite{Bozek:2015bha}. 
Local fluctuations of the initial flow increase the decorrelation of the final flow \cite{Pang:2015zrq}. Finally, dynamical hydrodynamic fluctuations \cite{Kapusta:2011gt} lead to fluctuations of the final flow of a finite range in rapidity \cite{Sakai:2018sxp}. Also thermodynamic fluctuation and contributions from  energy deposition from jets could contribute to the flow decorrelation. The contribution to the local collective flow from initial momentum flow, hydrodynamic fluctuations, thermal fluctuations, or jets is expected to be, in the first approximation, independent of the average collective flow. Therefore, it could be effectively described as a random component added to the flow vector as well.

%%%%%%%%%%%%%%%%%%%%%%%%%%%%%%%%%%%%%%%%%%%%%%%%%%%%%%%%%
\section{Factorization breaking coefficient of the collective flow}

\label{sec:facbreak}

The factorization breaking coefficient of the collective flow, which is equivalent to the correlation coefficient for the complex variables $(V_n(\eta))^m$ and $(V_n(-\eta))^m$, can be generalized to any power of the flow:
\begin{widetext}
\begin{equation}
  \rho_{V_n}^{(m)}(\eta,-\eta)= 
  \frac{\langle (V(\eta))^m (V^\star(-\eta))^m\rangle}{\sqrt{\langle (V(\eta))^m (V^\star(\eta))^m \rangle\langle (V(-\eta))^m (V^\star(-\eta))^m \rangle}} \ . 
\label{eq:fac2bin}
\end{equation}
The factorization breaking coefficient of the $m$th power of the flow vector in Eq.~(\ref{eq:fac2bin}) can be estimated in principle from an average of the moments of the experimentally measured $q$ vectors, 
\begin{equation}
\rho_{V_n}^{(m)}(\eta,-\eta)=\frac{\langle (q(\eta))^m (q^\star(-\eta))^m\rangle}{\sqrt{\langle (q(\eta))^m (q^\star(\eta))^m \rangle\langle (q(-\eta))^m (q^\star(-\eta))^m \rangle}} \ .
\end{equation}
\end{widetext}
However, for correlations in pseudorapidity the measure is strongly influenced by nonflow effects \cite{Bozek:2010vz}. In the models results shown here, the nonflow correlations are reduced by oversampling the final state hadrons for each hydrodynamic evolution event. We chose  the number of oversampling so that  the remaining contribution of nonflow resonance decays to the relative error is  smaller than the statistical errors in the calculation. On the other hand, such observables can be used as an experimental estimate for higher moment factorization breaking coefficients in transverse momentum \cite{Bozek:2018nne, Bozek:2021mov, ALICE:2022dtx},  if rapidity gaps are used to reduce nonflow correlations. Similar relations could be studied for analogous higher order factorization breaking coefficients in transverse momentum.

\begin{widetext}
	
	\begin{figure}[th!]
		\begin{center}
			\begin{tabular}{cc}
				\includegraphics[scale=0.4]{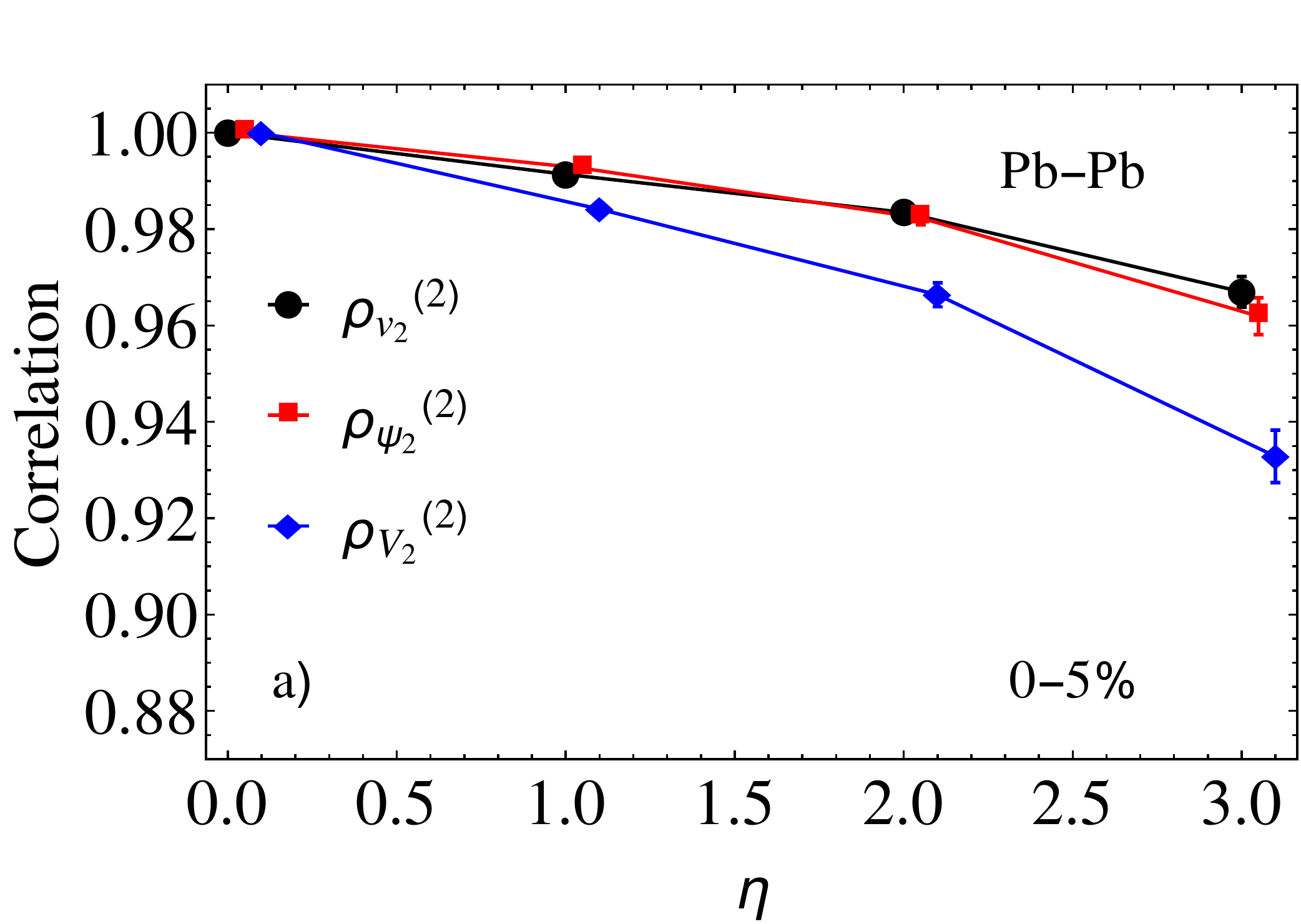} &
				\includegraphics[scale=0.4]{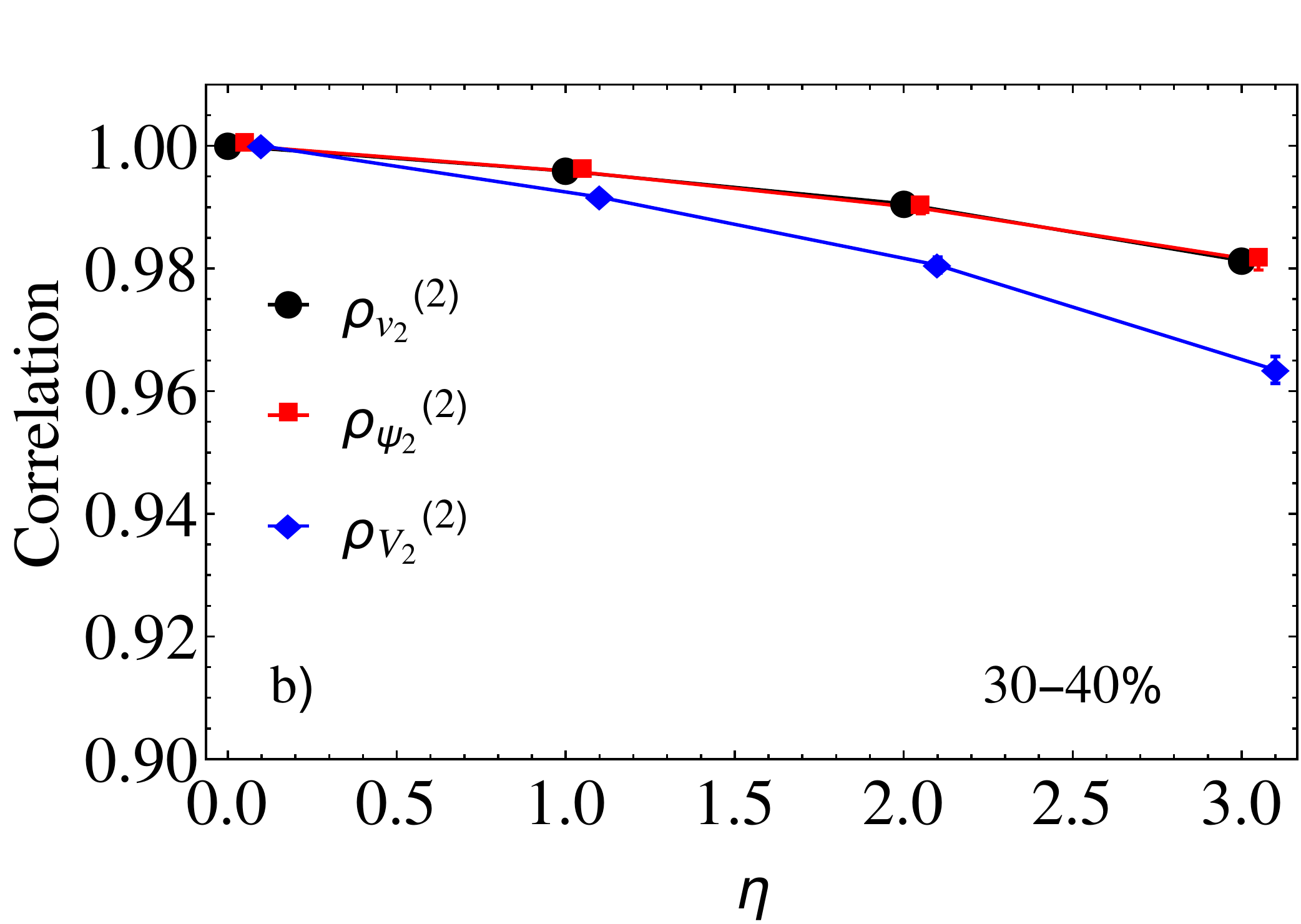} \\
				\includegraphics[scale=0.4]{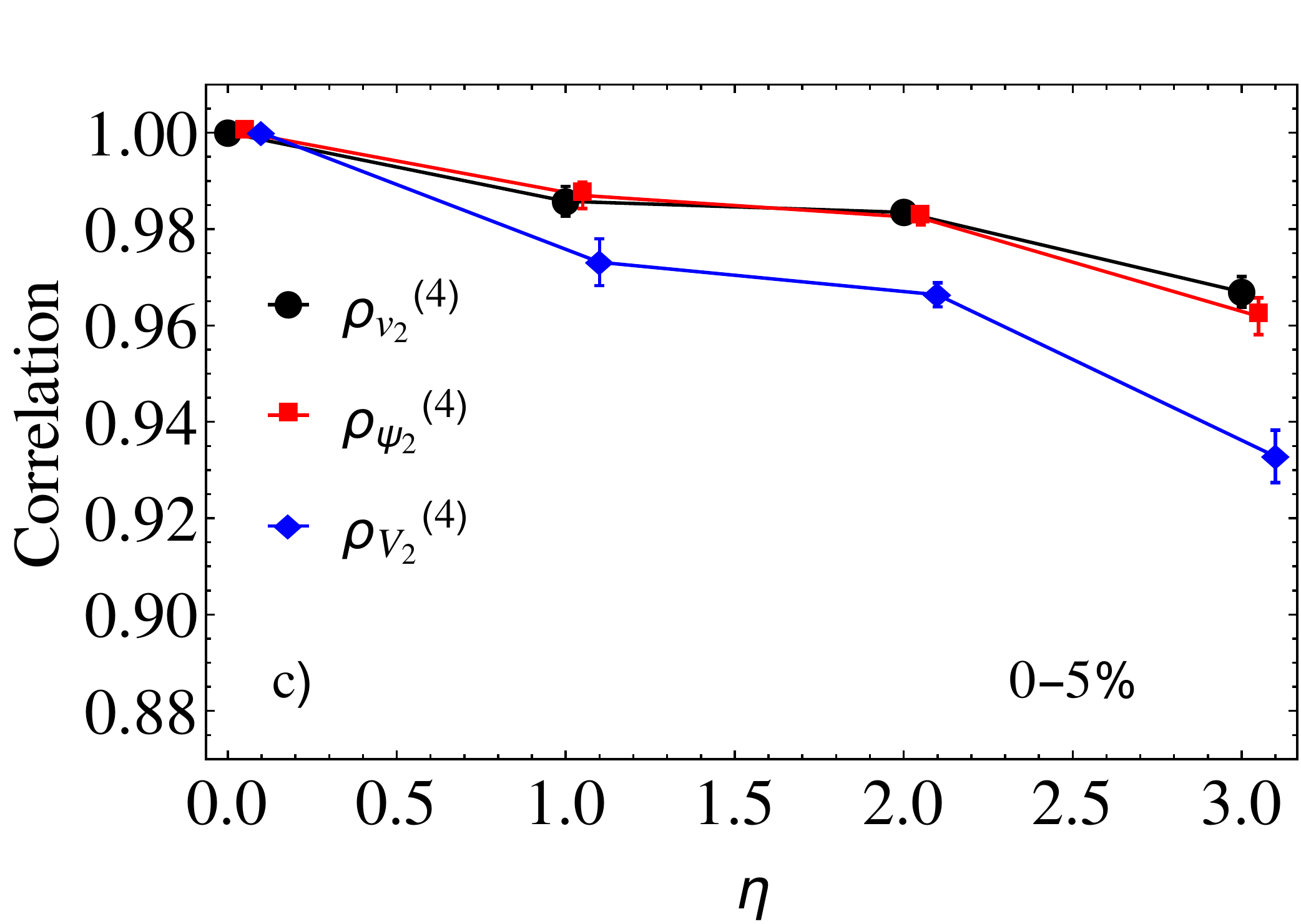} &
				\includegraphics[scale=0.4]{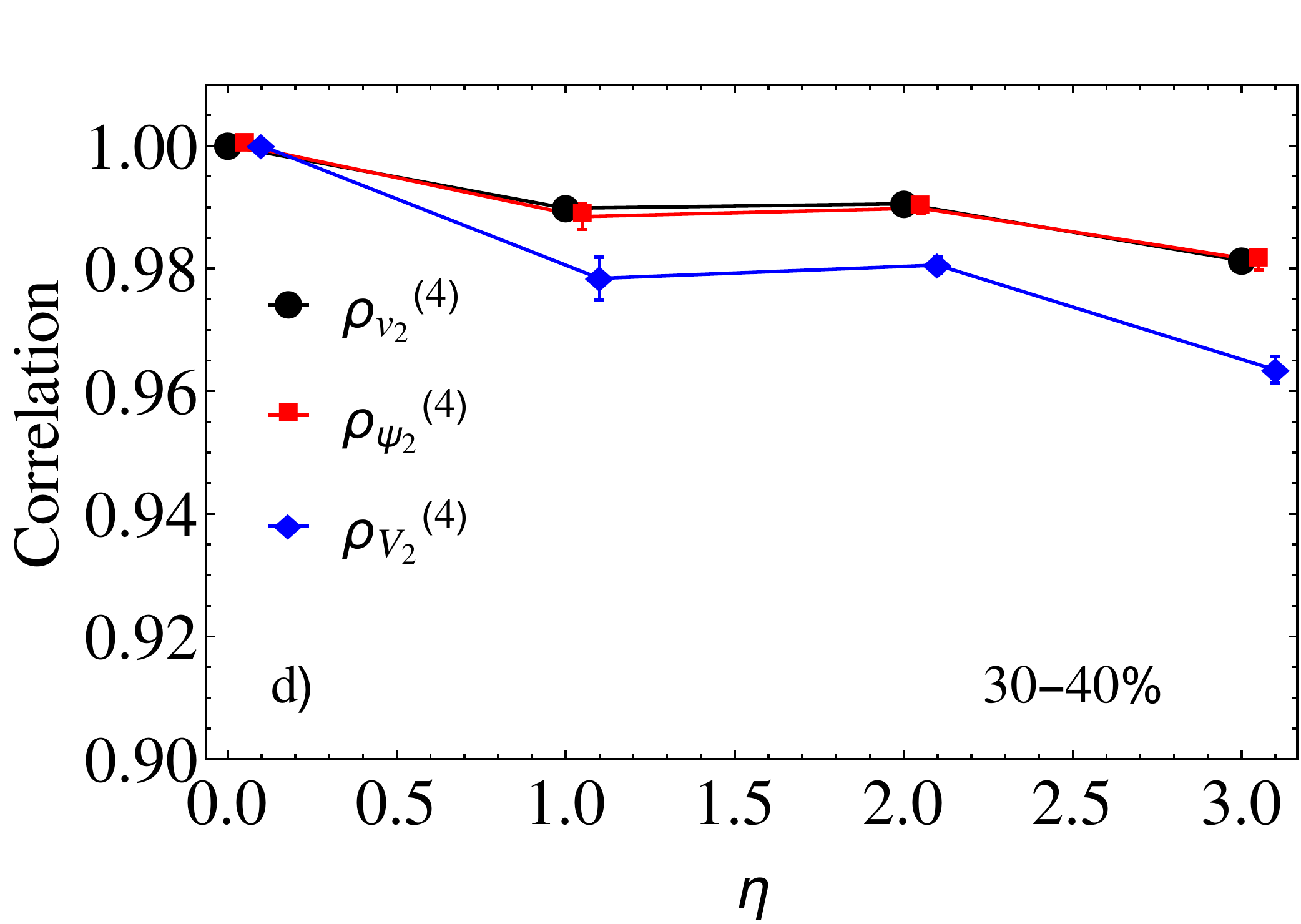} 		
			\end{tabular}		
			\caption{A comparison of the factorization breaking coefficient for the second power of flow magnitude (black lines, dots) with the factorization breaking coefficient for the second power of flow vectors (blue lines, diamonds) is shown in the top panels [central and semicentral collisions, panels (a) and (b) respectively].  The red lines with squares represents the flow angle factorization breaking coefficient, Eq.~(\ref{eq:fac2ang}). The factorization breaking coefficients for  the fourth power of the flow are shown in panels  (c) and (d).  All results are obtained from the viscous hydrodynamic model for Pb-Pb collisions at $\sqrt{s_{NN}}=5.02$TeV. } 
			\label{fig:corrangmagv2}
		\end{center}
	\end{figure}

	\begin{figure}[th!]
		\begin{center}
			\begin{tabular}{cc}
				\includegraphics[scale=0.4]{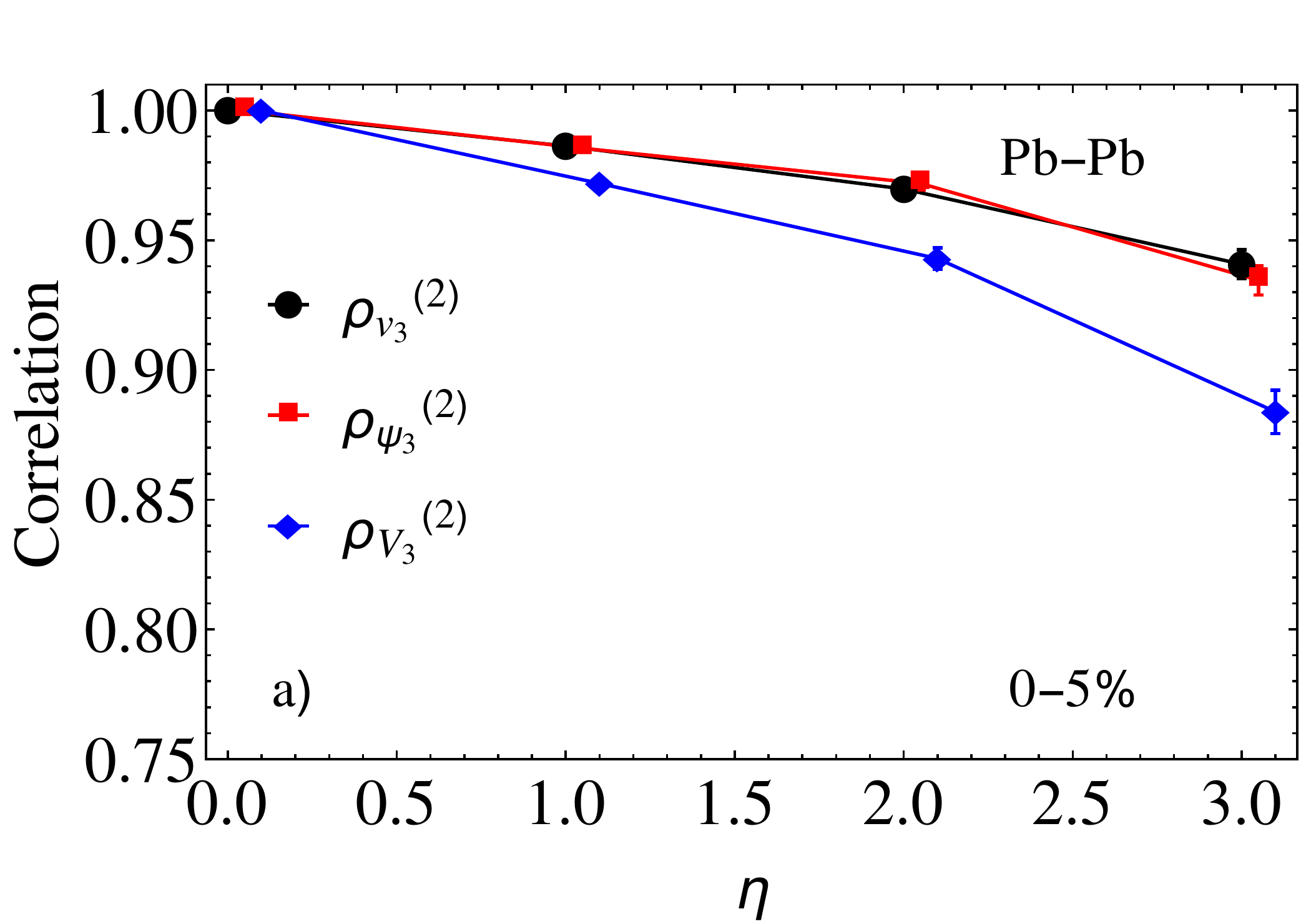} &
				\includegraphics[scale=0.4]{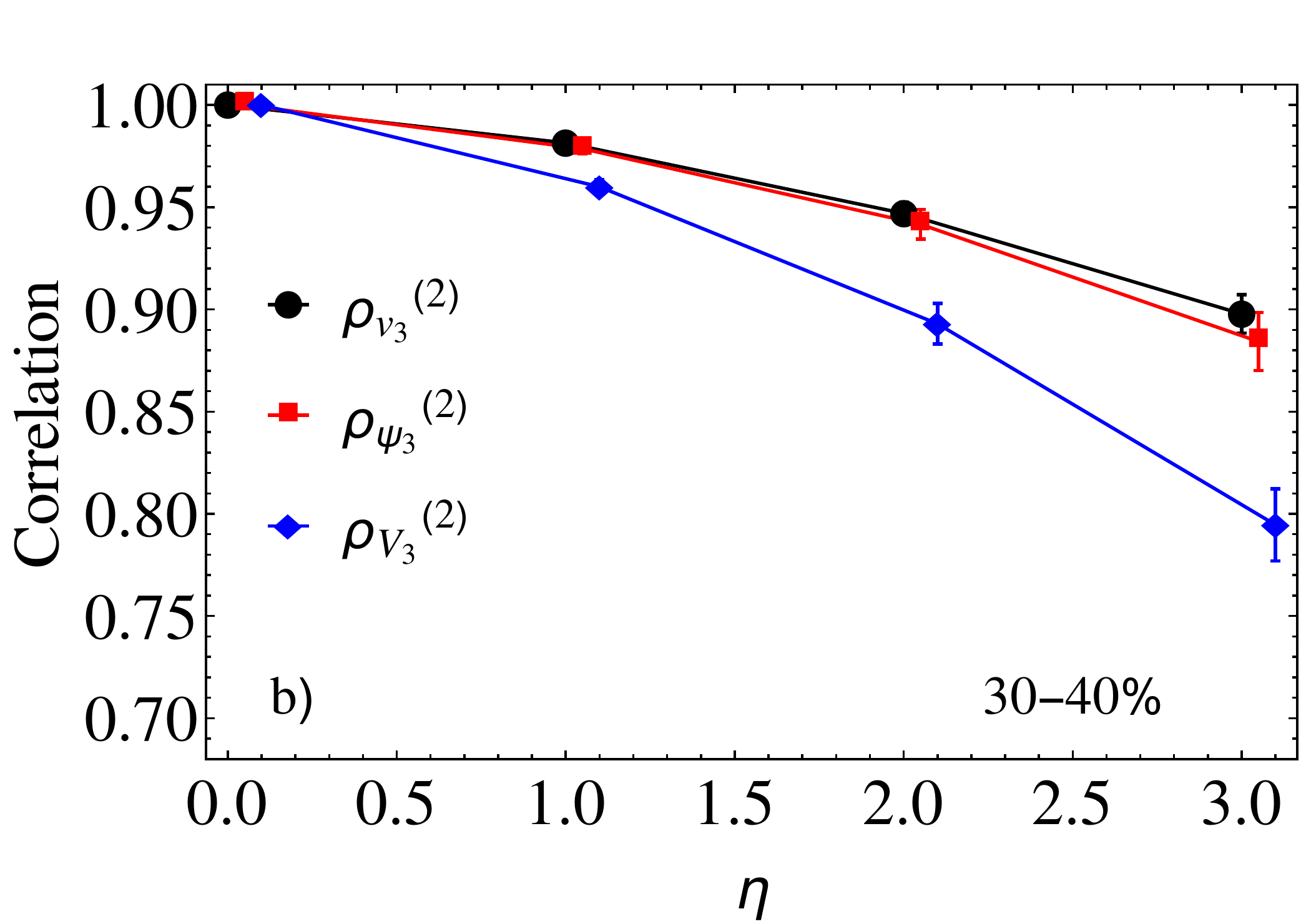} \\
				\includegraphics[scale=0.4]{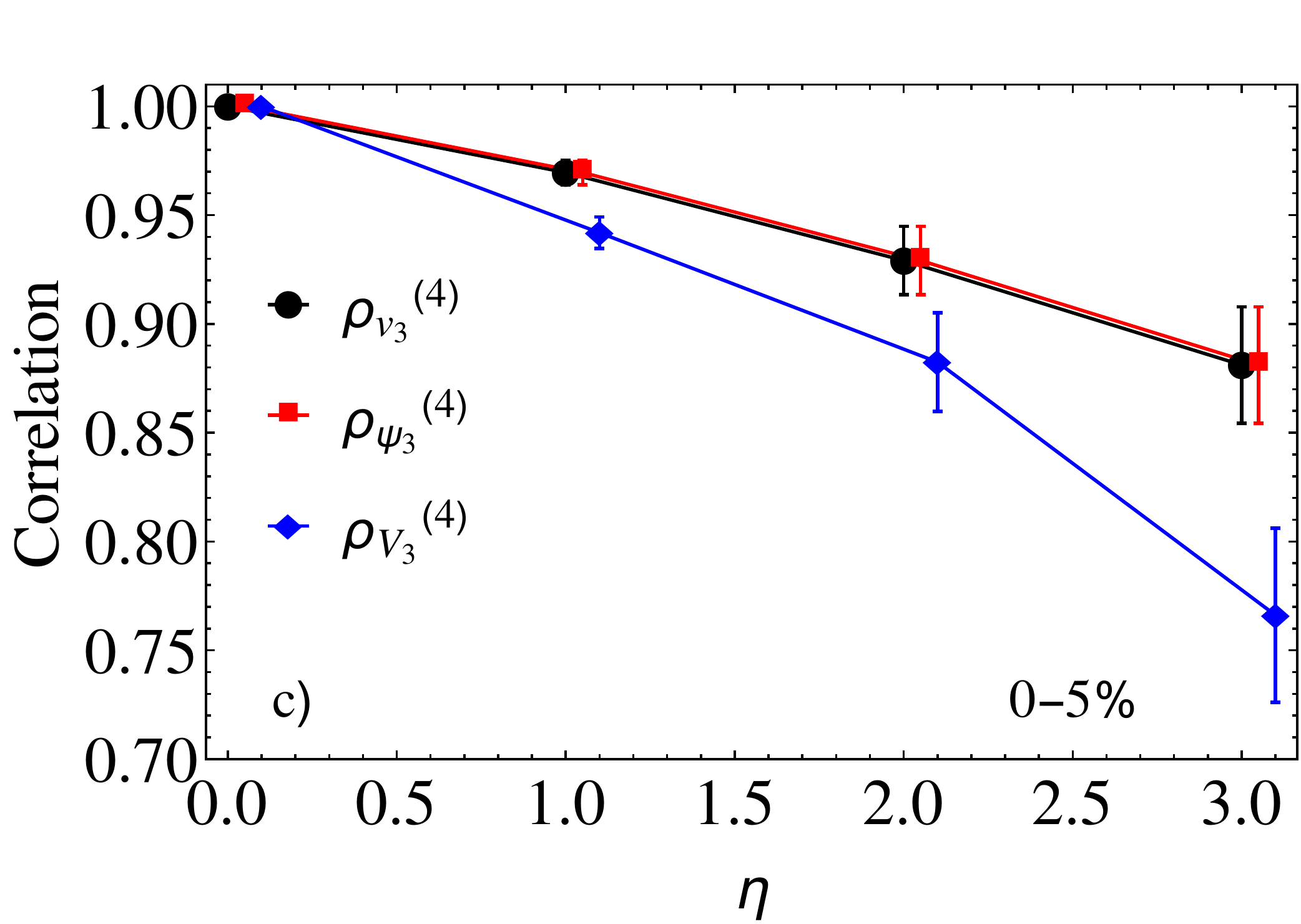} &
				\includegraphics[scale=0.4]{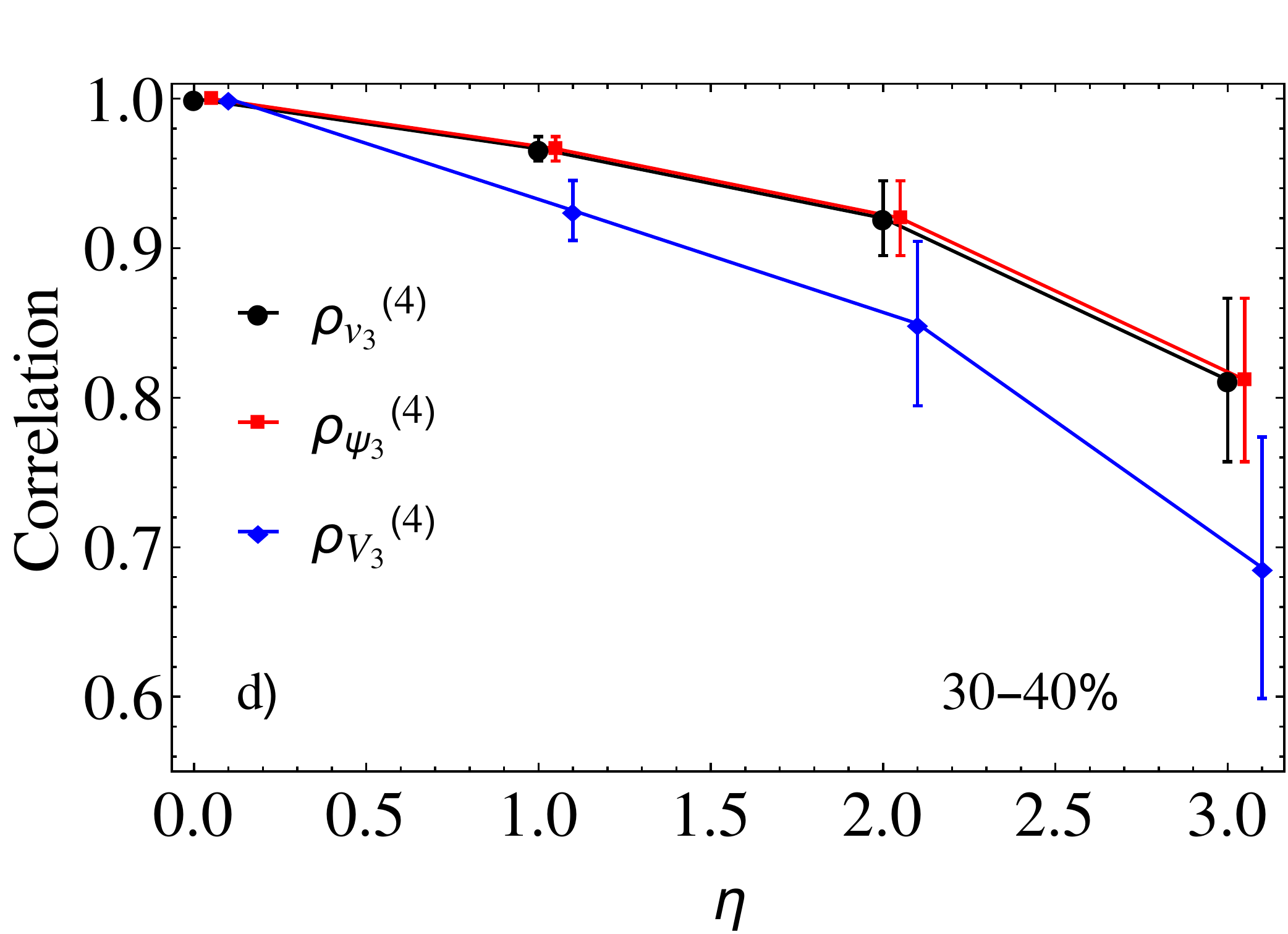} 		
			\end{tabular}		
			\caption{Same as in Fig.~\ref{fig:corrangmagv2} but for the triangular flow.} 
			\label{fig:corrangmagv3}
		\end{center}
	\end{figure}
	
\end{widetext}

In the random model of flow decorrelation, the flow in each bin is decomposed into  global and local random components. The factorization breaking coefficient takes the form
\begin{widetext}
	\begin{eqnarray}
	\rho_{V_n}^{(m)}(\eta,-\eta) & =& \frac{\langle (V_n+A_n +\Delta_n)^m(V_n^\star+A_n^\star-\Delta_n^\star)^m\rangle }{\sqrt{\langle (V_n+A_n +\Delta_n)^m(V_n^\star+A_n^\star+\Delta_n^\star)^m\rangle \langle (V_n+A_n -\Delta_n)^m(V_n^\star+A_n^\star-\Delta_n^\star)^m\rangle }} \nonumber \\
	& =& \frac{\langle (V_n^{'} +\Delta_n)^m(V_n^{'\star}-\Delta_n^\star)^m\rangle }{\langle (V_n^{'} +\Delta_n)^m(V_n^{'\star}+\Delta_n^\star)^m\rangle } \ ,
	\end{eqnarray}
\end{widetext}
where $V_n^{'}=V_n+A_n$. If the random component of the flow is relatively small $\delta_n =|\Delta_n|\ll |V_n|$ (then also $\delta_n =|\Delta_n|\ll |V_n^{'}|$), the factorization breaking coefficient can be expanded to second order in $\Delta$,
\begin{equation}
\rho_{V_n}^{(m)}(\eta,-\eta) \simeq 1 - 2 m^2 \frac{\langle v_n^{2m-2} \delta^2\rangle}{\langle v_n^{2m}\rangle} \ .
\label{eq:corrflowexp}
\end{equation}
The above formula shows the generic properties of the factorization-breaking coefficient. On general grounds, one expects that the lowest order dependence of $\Delta$ on the bin separation $\Delta\eta$ is linear \cite{Bozek:2010vz, Bozek:2015bna}. This leads to a quadratic dependence of the factorization-breaking coefficient on the bin separation,
\begin{equation}
\rho_{V_n}^{(m)}(\eta_1,\eta_2) \simeq 1 - 2 m^2 \kappa (\eta_1-\eta_2)^2 \ .
\end{equation}
The flow moment decorrelation, i.e. the deviation of the factorization breaking coefficient from unity, increases with the rank $m$ of the flow moment. The formula can be further simplified if the factorization of moments $v$ and $\delta$ is assumed:
\begin{eqnarray}
\rho_{V_n}^{(m)}(\eta,-\eta) &\simeq & 1 - 2 m^2 \frac{\langle v_n^{2m-2}\rangle\langle \delta^2\rangle}{\langle v_n^{2m}\rangle} \nonumber \\
&\simeq& 1- 2 m \frac{\langle \delta^2\rangle}{\langle v_n^{2}\rangle} \ ,\label{eq:corrflowexpsimple}
\end{eqnarray}
where the last equality is obtained for fluctuation-dominated flow. If the collective flow is dominated by fluctuations in the initial state, as for the triangular flow or the elliptic flow in central collisions, we have $\langle v_n^{2m} \rangle = m! \langle v_n^2\rangle^m$.

The deviation of the factorization breaking coefficient from $1$ comes from two effects, the flow angle decorrelation and flow magnitude decorrelation, as noted in \cite{Heinz:2013bua, Jia:2017kdq}. Both effects can be studied separately in model calculations \cite{Bozek:2018nne}. The factorization breaking coefficient for the flow magnitudes is:
\begin{equation}
\rho_{v_n}^{(m)}(\eta,-\eta) = \frac{\langle v_n(\eta)^m v_n(-\eta)^m \rangle}
{\sqrt{\langle v_n(\eta)^{2m} \rangle\langle v_n(-\eta)^{2m} \rangle}} \ .\label{eq:fac2mag}
\end{equation}
The factorization breaking coefficient for the flow magnitudes could be estimated experimentally (at least in principle) for {\it even} $m=2k$:
\begin{equation}
\rho_{v_n}^{(m)}(\eta,-\eta) = \frac{\langle (q_n(\eta))^k (q_n^\star(\eta))^k (q_n(-\eta))^k (q_n^\star(-\eta))^k \rangle}
{\sqrt{\langle (q_n(\eta))^{2k} (q_n^\star(\eta))^{2k} \rangle\langle (q_n(\eta))^{2k} (q_n^\star(\eta))^{2k} \rangle}} \ .
\end{equation}

\begin{figure}[th!]
	%\vskip 3mm
	\includegraphics[scale=0.4]{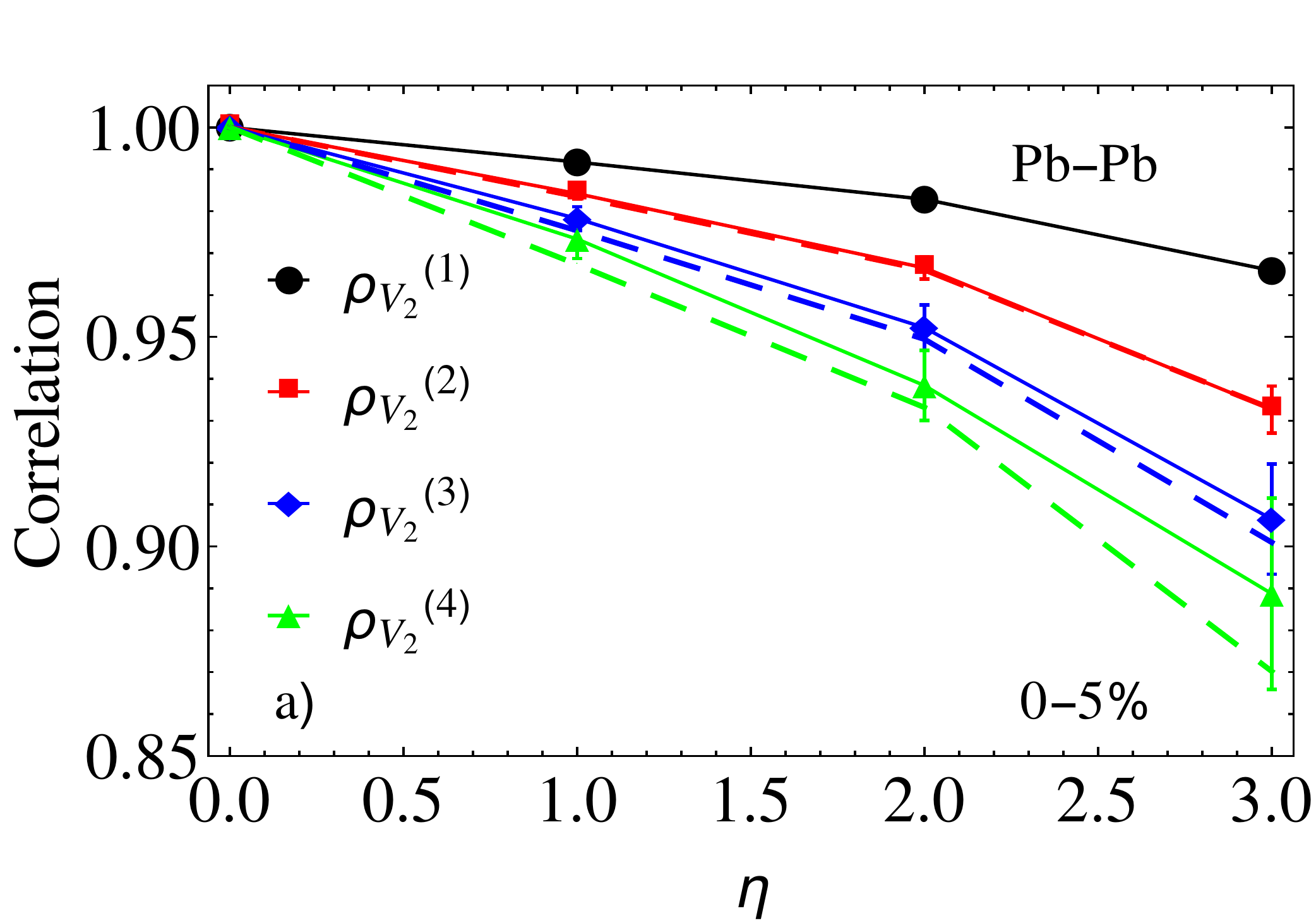} \\
	\includegraphics[scale=0.4]{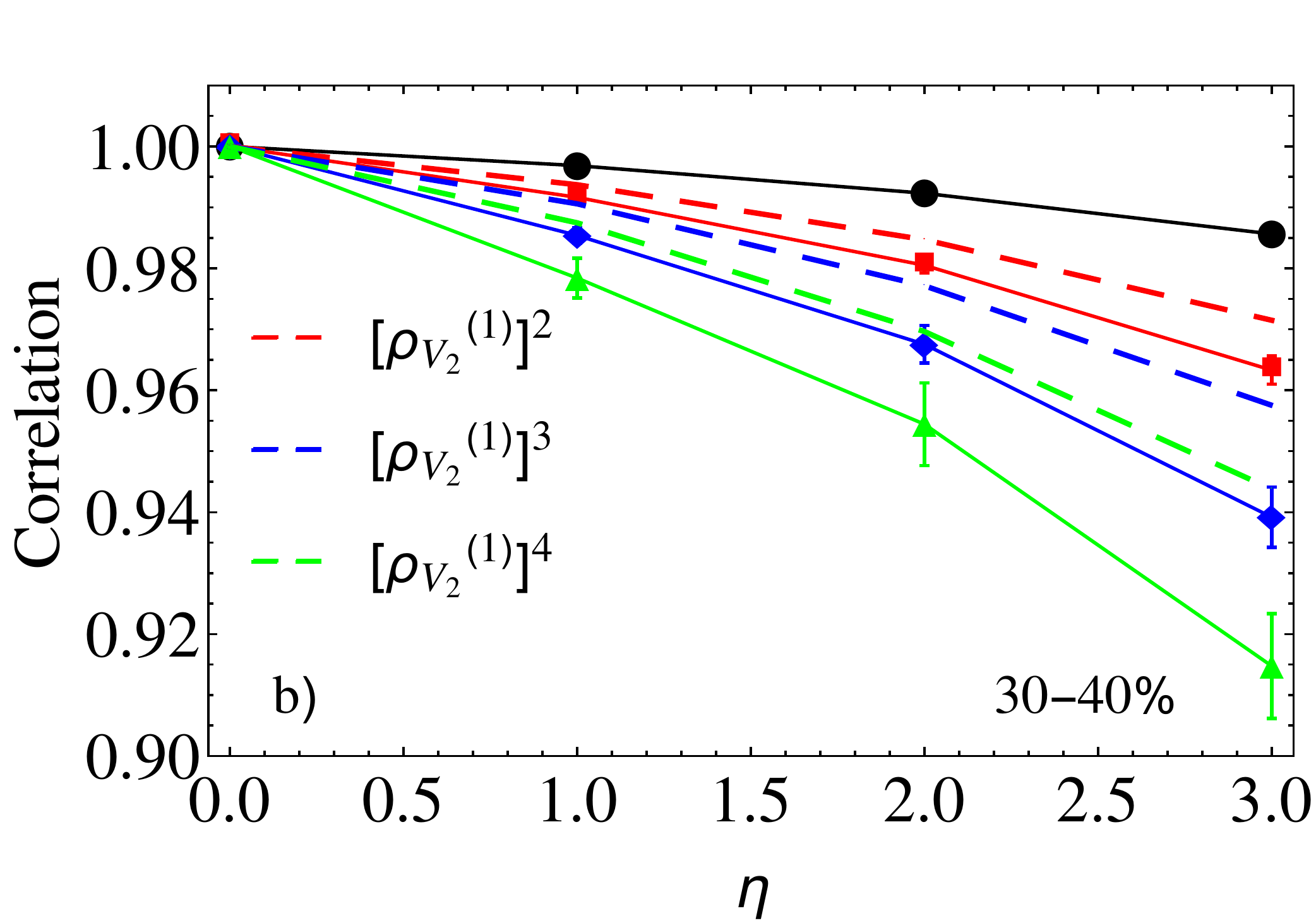}
	\caption{Flow factorization breaking coefficients for different moments of flow vectors calculated in the hydrodynamic model. Panels (a) and (b) are for central  ($0-5$\%) and semi-central ($30-40$\%) collisions, respectively. The factorization breaking coefficients $\rho^{(1)}$, $\rho^{(2)}$, $\rho^{(3)}$, and $\rho^{(4)}$ are denoted with circles, squares, diamonds, and triangles respectively. The powers of the first order coefficient $[\rho^{(1)}]^m$ are represented with dashed lines of the same color as the corresponding coefficients $\rho^{(m)}$.} 
	\label{fig:facdiffmv2}
\end{figure}

\begin{figure}[th!]
	%\vskip 3mm
	\includegraphics[scale=0.4]{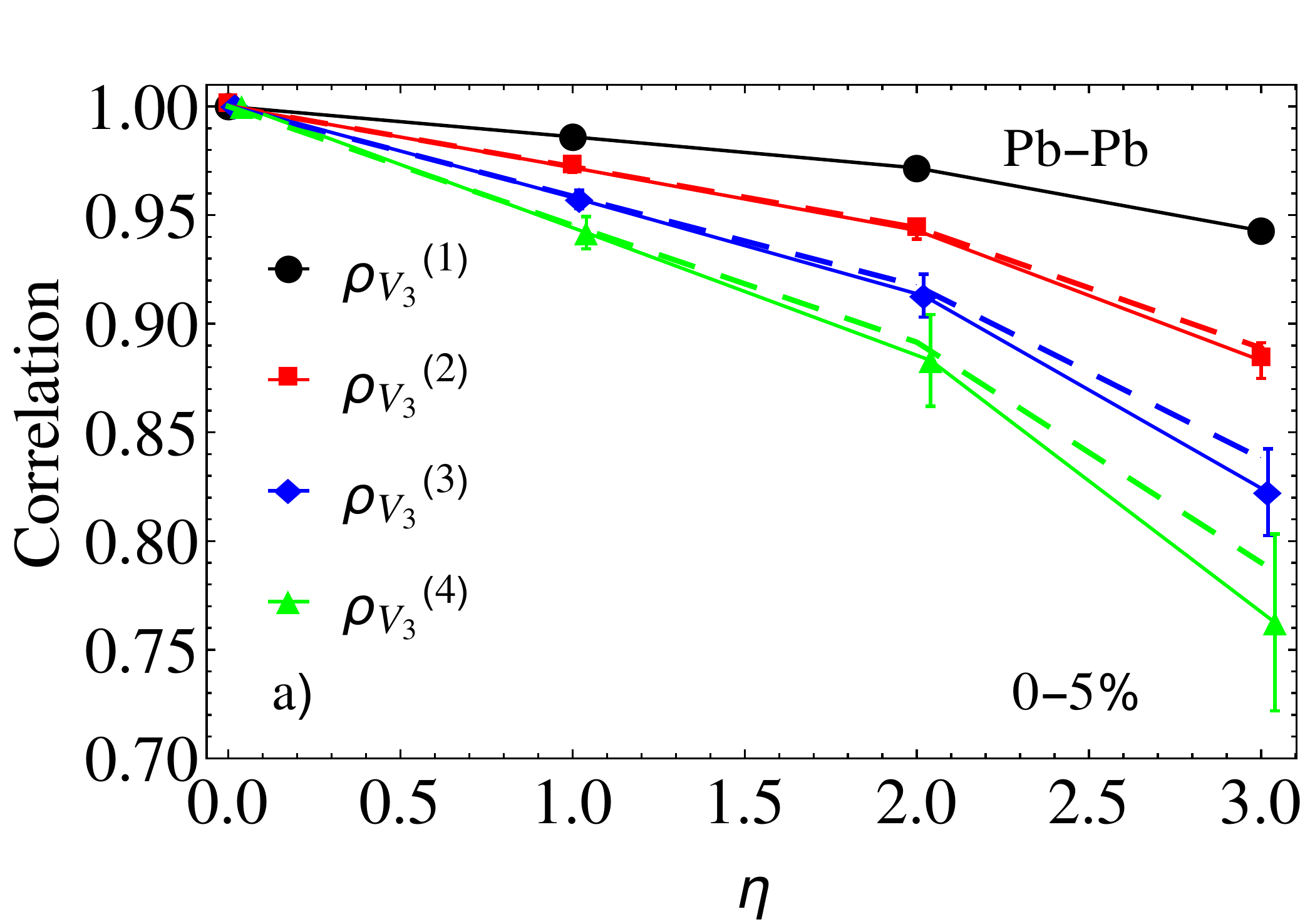} \\
	\includegraphics[scale=0.4]{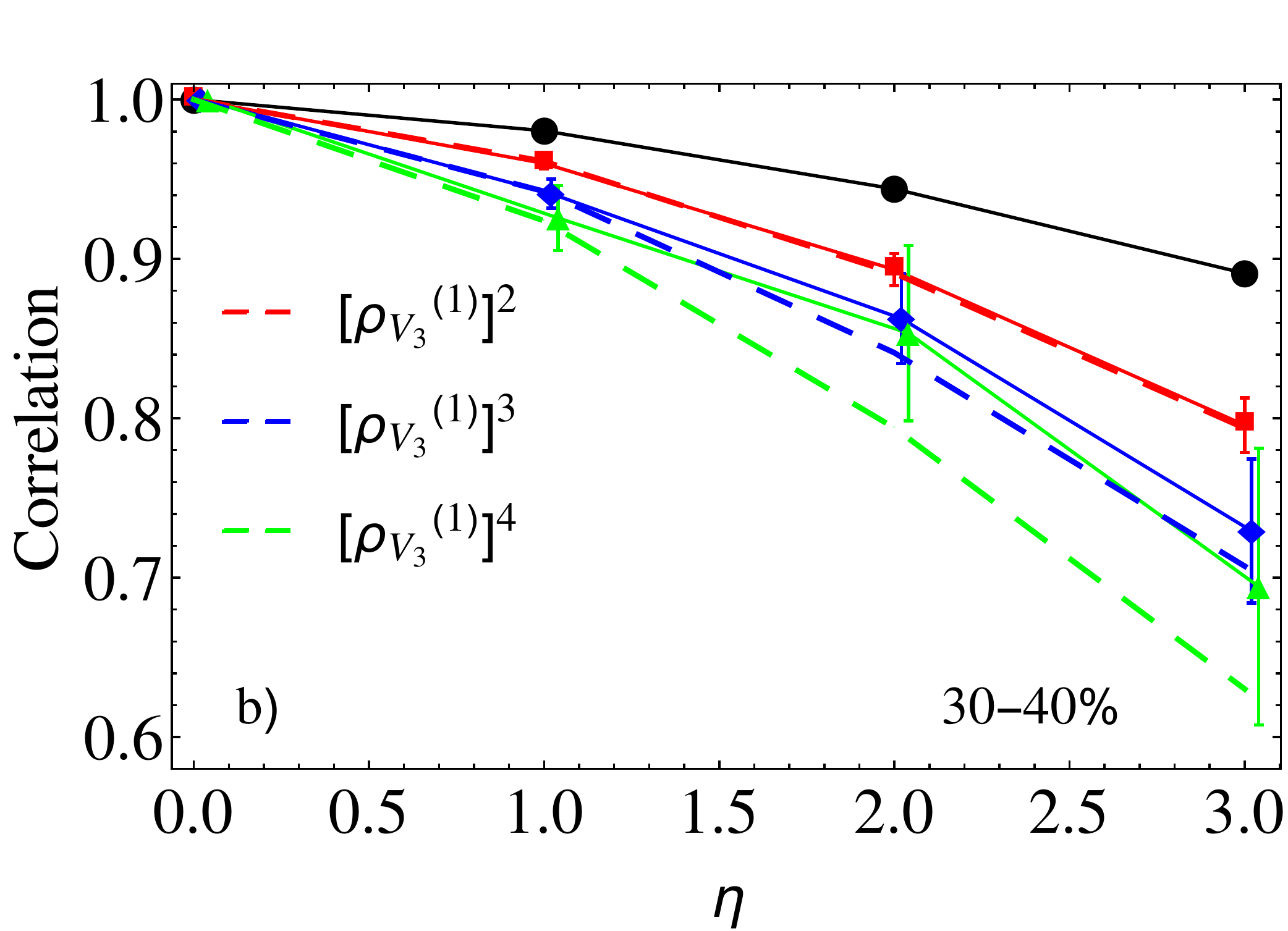}
	\caption{same as in Fig.~\ref{fig:facdiffmv2} but for the triangular flow.} 
	\label{fig:facdiffmv3}
\end{figure}

%\begin{widetext}
The flow magnitude factorization breaking can be  calculated using our model with a random local flow component,
\begin{eqnarray}
\rho_{v_n}^{(m)}(\eta,-\eta) & =& \frac{\langle |V_n+A_n +\Delta_n|^m|V_n+A_n-\Delta_n|^m\rangle }{\sqrt{\langle |V_n+A_n +\Delta_n|^{2m}\rangle \langle |V_n+A_n -\Delta_n|^{2m}\rangle }} \nonumber \\ 
& =& \frac{\langle |V_n^{'} +\Delta_n|^m|V_n^{'}-\Delta_n|^m\rangle }{\langle |V_n^{'} +\Delta_n|^{2m}\rangle } \ .
\end{eqnarray}
%\end{widetext}
Expansion to the second order in $\delta/v$ gives
\begin{equation}
\rho_{v_n}^{(m)}(\eta,-\eta) \simeq 1 -  m^2  \frac{\langle  v_n^{2m-2} \delta^2\rangle}{\langle v_n^{2m}\rangle} \ .
\end{equation}
In the random model of flow decorrelation, the decorrelation of the flow vector magnitudes is approximately one half of the decorrelation of the flow vectors. The same property can be observed in the viscous hydrodynamic model results (Figs. \ref{fig:corrangmagv2} and \ref{fig:corrangmagv3}). 

To estimate the flow angle decorrelation between two bins centered at $\eta$ and $-\eta$ a simple average of the cosine of flow angle difference could be used:
\begin{equation}
\left\langle \cos\left(m n (\Psi_n(\eta)-\Psi_n(-\eta))\right) \right\rangle =\left\langle \frac{\left(V_n(\eta)V_n^\star(-\eta)\right)^m}{\left(|V_n(\eta)||V_n(-\eta)|\right)^m}\right\rangle \ .
\label{eq:angle}
\end{equation}
The flow angle decorrelation defined above cannot be directly measured in the experiment, unlike the flow vector factorization breaking coefficient in Eq.~(\ref{eq:fac2bin}) (for any $m$) or the flow magnitude factorization breaking coefficient in Eq.~(\ref{eq:fac2mag}) (for even $m$). Note that in the definition of flow factorization breaking the event average is taken separately in the numerator and the denominator. In contrast, for the angle decorrelation defined in Eq.~(\ref{eq:angle}), the event average is taken for the whole ratio. The expansion of the flow angle decorrelation for small $\delta/v$ gives
\begin{equation}
\langle \cos\left(m n (\Psi_n(\eta)-\Psi_n(-\eta))\right) \rangle \simeq 
1- m^2 \left\langle\frac{\delta_n^2}{v_n^2} \right\rangle.
\label{eq:angleimag}
\end{equation}
The simple flow angle decorrelation of order $m$ in Eq.~(\ref{eq:angle}) should not be confused with the angular component of the flow factorization breaking, $\rho^{(m)}_{V_n}(\eta,-\eta)$, Eq.~(\ref{eq:fac2bin}). Under the event average, the flow magnitude to power $m$ is present in the flow factorization breaking coefficient of order $m$. It has been noticed that the flow angle decorrelation is strongly anticorrelated with flow magnitude in the event \cite{Bozek:2017qir}. Therefore, the relevant flow angle factorization breaking coefficient  in order $m$ is defined as
\begin{equation}
\rho_{\Psi_n}^{(m)}(\eta,-\eta)= \frac{\langle v^{2m} \cos\left(m n (\Psi_n(\eta)-\Psi_n(-\eta))\right) \rangle }{\langle v^{2m} \rangle} \ .
\label{eq:fac2ang}
\end{equation}
In experiment, the flow angle factorization breaking coefficient can be estimated as  the ratio of the flow vector and flow vector magnitude factorization breaking coefficients,
\begin{equation}
\rho_{\Psi_n}^{(m)}(\eta,-\eta)=\frac{\rho_{V_n}^{(m)}(\eta,-\eta) }{\rho_{v_n}^{(m)}(\eta,-\eta) } \ , 
\end{equation}
with almost the same results as from the definition in Eq.~(\ref{eq:fac2ang}).
 For small flow decorrelation, we have
\begin{equation}
\rho_{\Psi_n}^{(m)}(\eta,-\eta) \simeq 1 - m^2 \frac{\langle v^{2m-2}\delta^2\rangle }{\langle v^{2m} \rangle}.
\end{equation}

In Figs.~\ref{fig:corrangmagv2} and \ref{fig:corrangmagv3}, we compare the results for  measures of flow decorrelation in two pseudorapidity bins: the flow factorization breaking coefficient, the flow magnitude factorization breaking coefficient, and the flow angle factorization breaking coefficient  in the hydrodynamic model. The numerical results follow the relation obtained in our random flow decorrelation model. In particular, the flow factorization breaking can be decomposed into the flow magnitude factorization breaking and the flow angle decorrelation,
\begin{equation}
\rho_{V_n}^{(m)}(\eta,-\eta)\simeq \rho_{v_n}^{(m)}(\eta,-\eta) \rho_{\Psi_n}^{(m)}(\eta,-\eta), 
\end{equation}
where, with considerable accuracy, we find:
\begin{equation}\label{eq:acc}
[1-\rho_{\Psi_n}^{(m)}(\eta,-\eta)]\simeq[1- \rho_{v_n}^{(m)}(\eta,-\eta)] \simeq[1-\rho_{V_n}^{(m)}(\eta,-\eta)]/2.
\end{equation}
The flow angle decorrelation is roughly the same as the flow magnitude decorrelation (compare red and black lines in Figs. \ref{fig:corrangmagv2} and \ref{fig:corrangmagv3}). If the above relation holds, Eq~(\ref{eq:acc}) could be used as an experimental estimate of the flow angle decorrelation also for odd $m$.

Finally, we test the relation given in Eq.~(\ref {eq:corrflowexpsimple}) between flow factorization coefficients of different order $m$. When the event-by-event flow follows the Bessel-Gaussian distribution \cite{Mehrabpour:2020wlu,Mehrabpour:2023ign} (central collisions for the elliptic flow and all centralities for the triangular flow) and for small deviations of the factorization breaking coefficients from $1$, one finds: 
\begin{equation}
\rho_{V_n}^{(m)}(\eta,-\eta)\approx\Big[ \rho_{V_n}^{(1)}(\eta,-\eta) \Big]^m \ .
\label{eq:app}
\end{equation}
Numerical results from the viscous hydrodynamic model confirm these relations as illustrated in Figs. \ref{fig:facdiffmv2} and \ref{fig:facdiffmv3}.
The factorization breaking coefficient of order $m=2,3$ can be approximated as a power of the first-order factorization breaking coefficient for the elliptic flow at $0$-$5$\% centrality and the triangular flow at both centralities studied. The elliptic flow in semicentral events ($30$-$40$\%) is not dominated by fluctuations and the relation from Eq.~(\ref {eq:corrflowexpsimple}) for the factorization breaking coefficients of different orders is not true. $\rho^{(4)}$ deviates strongly from $1$ and the approximate relation in Eq.~(\ref{eq:app}) is broken.

\section{Flow angle decorrelation and the overall flow magnitude}

\label{sec:flowangle}

\begin{figure}[th!]
	%\vskip 3mm
	\includegraphics[scale=0.4]{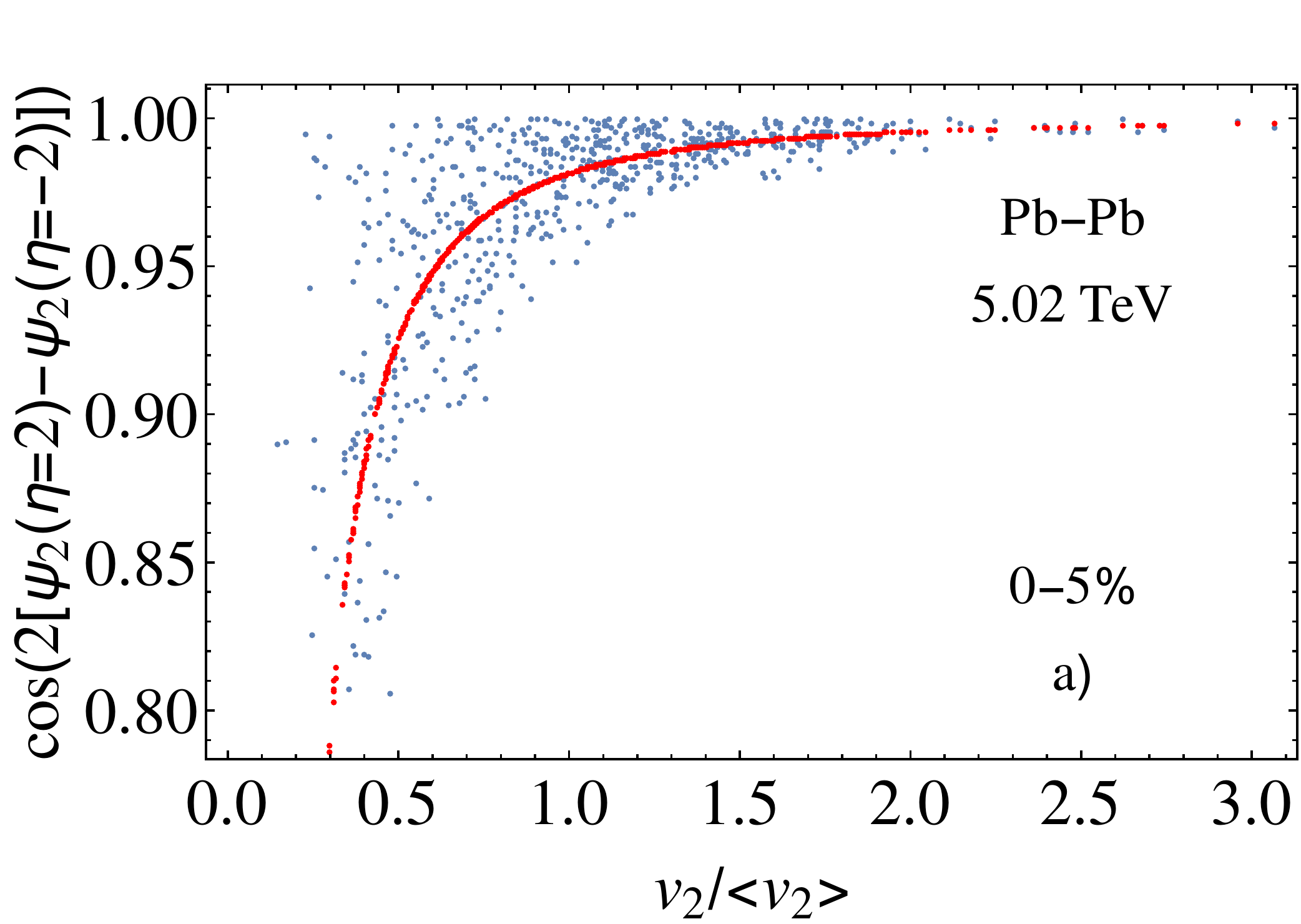} \\
	\includegraphics[scale=0.4]{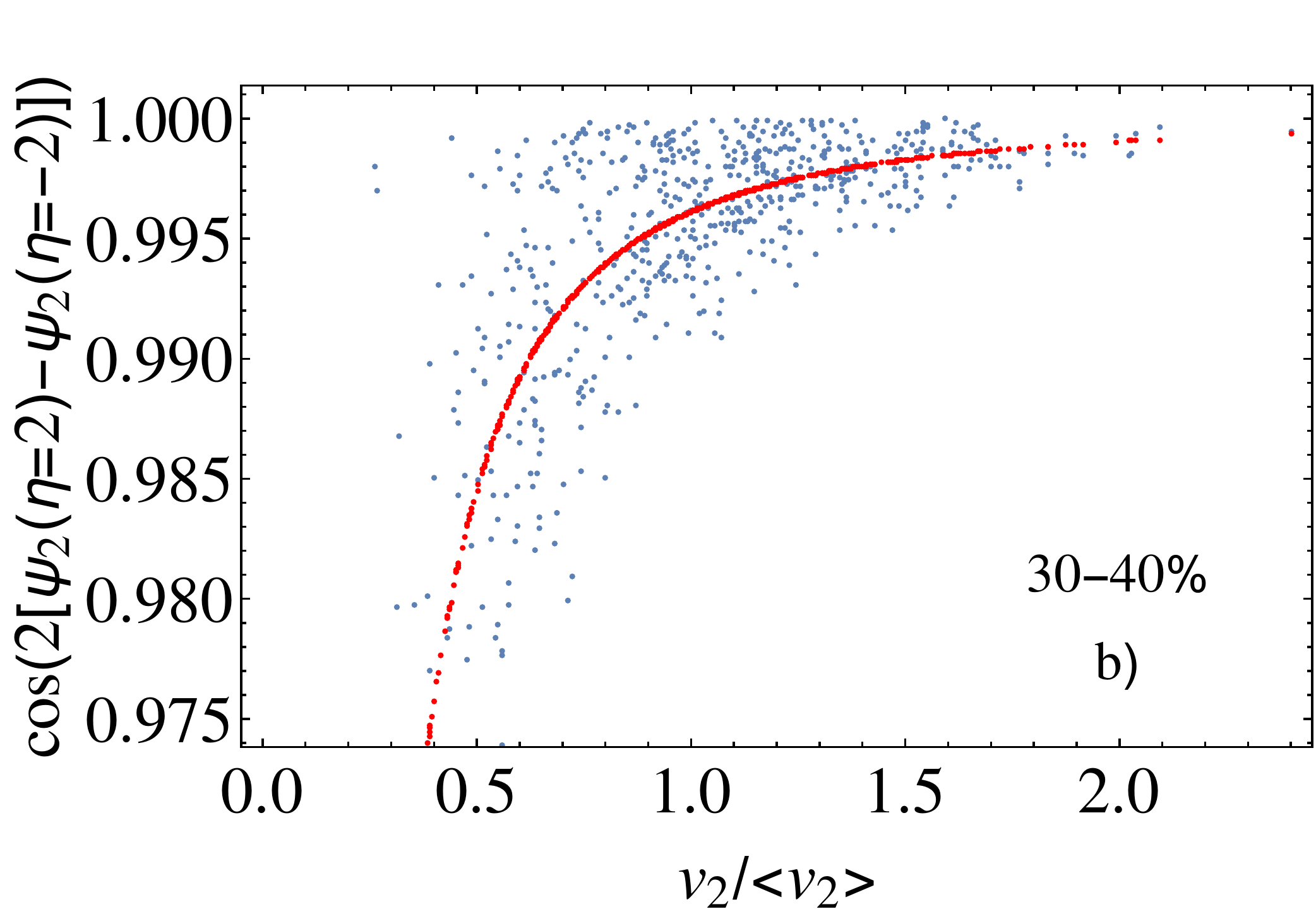}
	\caption{Scatter plot of the scaled elliptic flow in an event $v/\langle v \rangle$ versus $\cos(n(\Psi_n(\eta)-\Psi_n(-\eta))$ for central (panel a)) and semicentral (panel b)) collisions. All the points are from the viscous hydrodynamic model. The red points represent the expected flow angle decorrelation as a function of the fixed value of the flow in the event, Eq. (\ref{eq:expcos}).} 
	\label{fig:scatterv2}
\end{figure}

\begin{figure}[th!]
	%\vskip 3mm
	\includegraphics[scale=0.4]{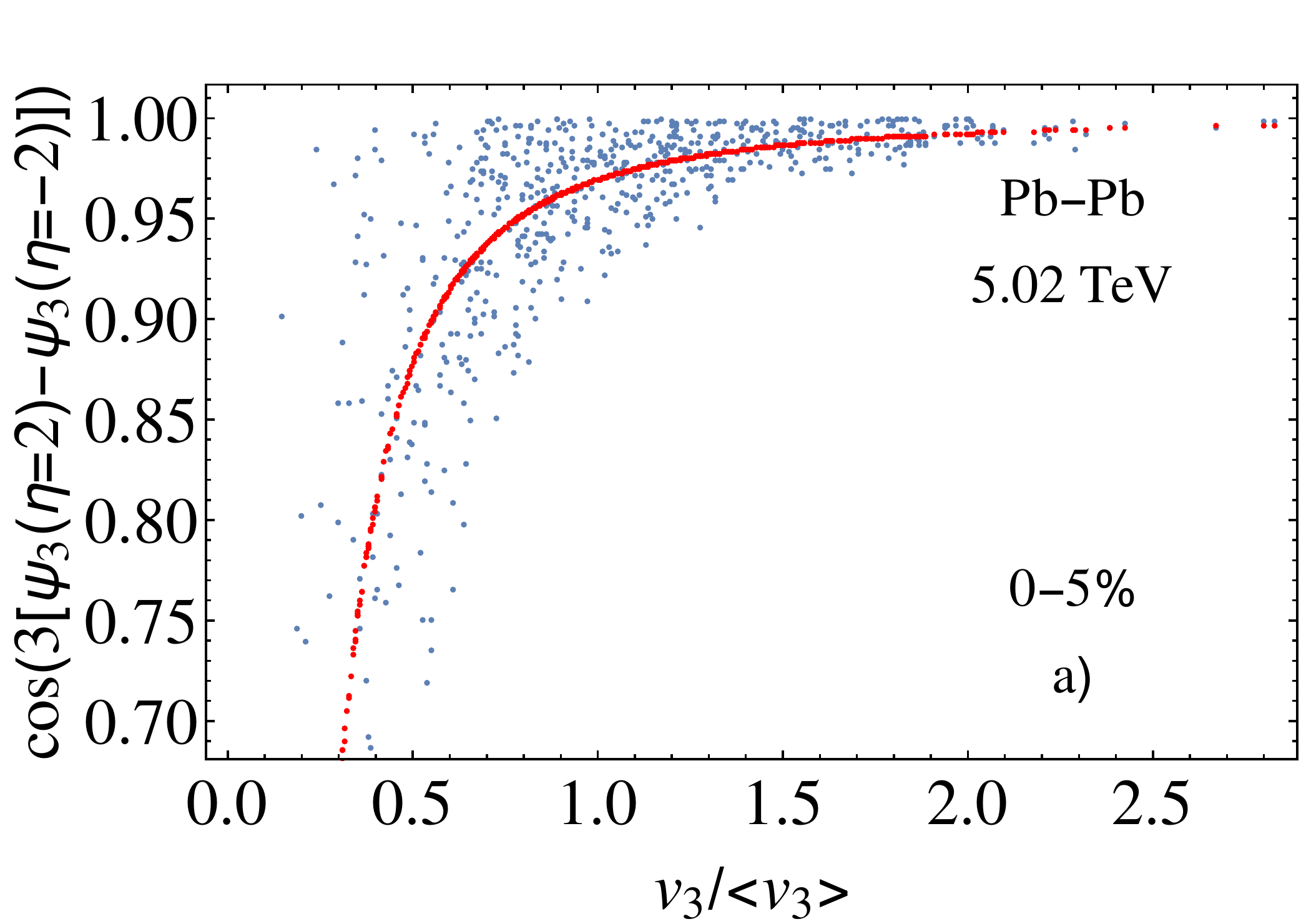} \\
	\includegraphics[scale=0.4]{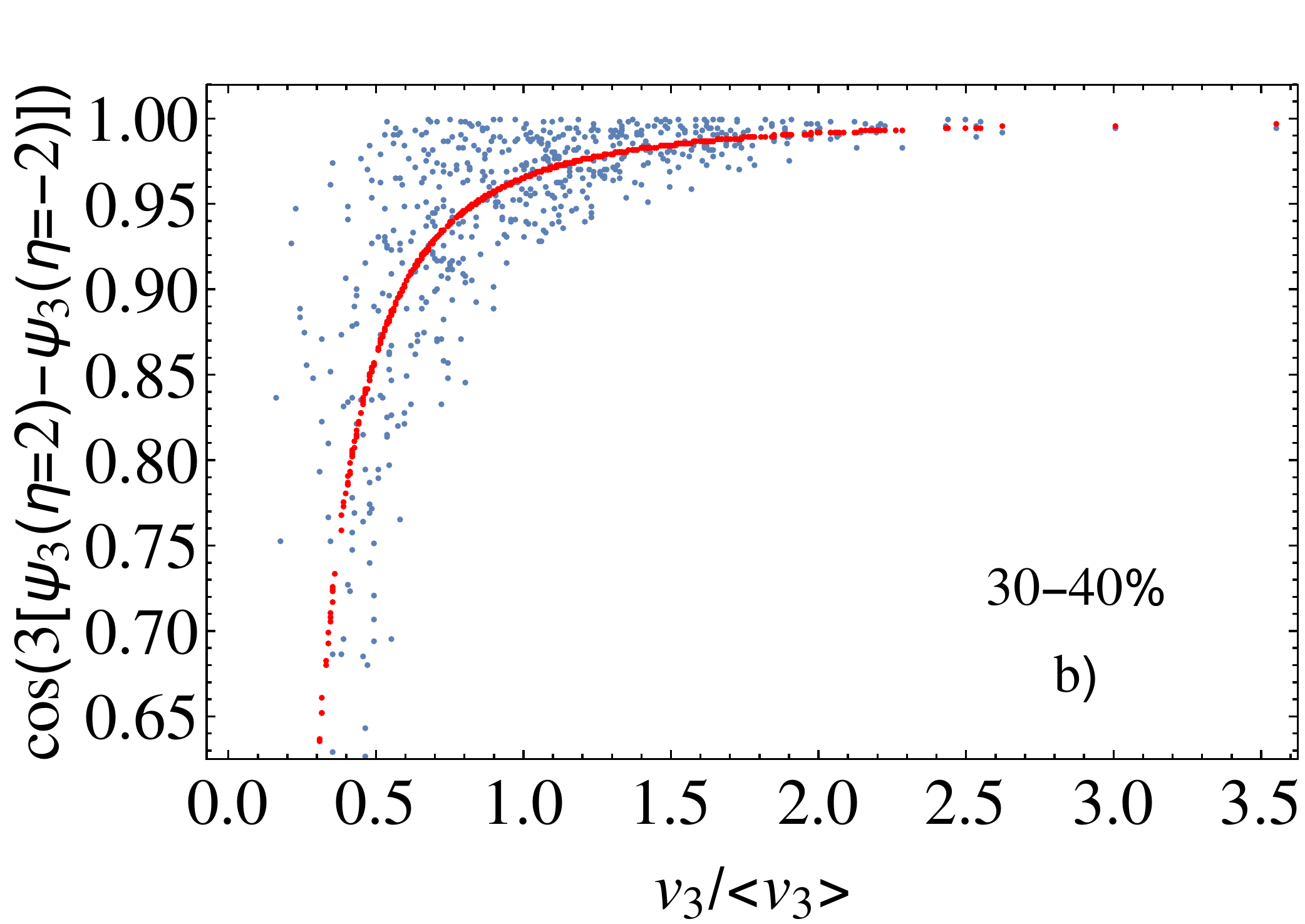}
	\caption{Same as in Fig.~\ref{fig:scatterv2} but for the triangular flow.} 
	\label{fig:scatterv3}
\end{figure}

The flow angle decorrelation is stronger if the overall flow is small. The largest flow decorrelation occurs for the triangular flow  and for the elliptic flow in central events \cite{Bozek:2010vz}. Moreover, for a given centrality class, in events with a smaller magnitude of the overall flow, the angle decorrelation is the largest \cite{Bozek:2017qir, Bozek:2018nne}. In Figs.~\ref{fig:scatterv2} and \ref{fig:scatterv3} are shown the scatter plots of the overall flow magnitude in the event and of the cosine of the flow decorrelation for a sample of central and semicentral events.
The increase of the flow angle decorrelation with decreasing overall flow magnitude $v$ can be understood from Eq.~(\ref{eq:angleimag}). Taking 
\begin{equation}
\langle \cos\left( n (\Psi_n(\eta)-\Psi_n(-\eta))\right) \rangle \simeq 
1- \left\langle  \frac{\delta_n^2}{v_n^2} \right\rangle,
\end{equation}
at a fixed value of the flow magnitude, $v_n$, one has:
\begin{equation}
\langle \cos\left( n (\Psi_n(\eta)-\Psi_n(-\eta))\right) \rangle|_{v_n} \simeq 
1-\frac{\langle \delta_n^2\rangle  }{v_n^2}.
\label{eq:expcos}
\end{equation}
We illustrate  the relation between the flow angle decorrelation and the  overall flow magnitude for $m=1$. 
The anticorrelation of the flow angle decorrelation $\Psi_n(\eta)-\Psi_n(-\eta)$ with $v_n$ [Eq. (\ref{eq:expcos}] is denoted with red points in Figs. \ref{fig:scatterv2} and \ref{fig:scatterv3}. The random model of flow decorrelation explains the anticorrelation between the overall flow magnitude $v_n$ and the angle correlation $\langle \cos\left( n (\Psi_n(\eta)-\Psi_n(-\eta))\right) \rangle$ observed in numerical simulations of the hydrodynamic model.

\section{three-bin and four-bin measures of flow decorrelation}

\label{sec:34bins}

\begin{figure}[th!]
	%\vskip 3mm
	\includegraphics[scale=0.4]{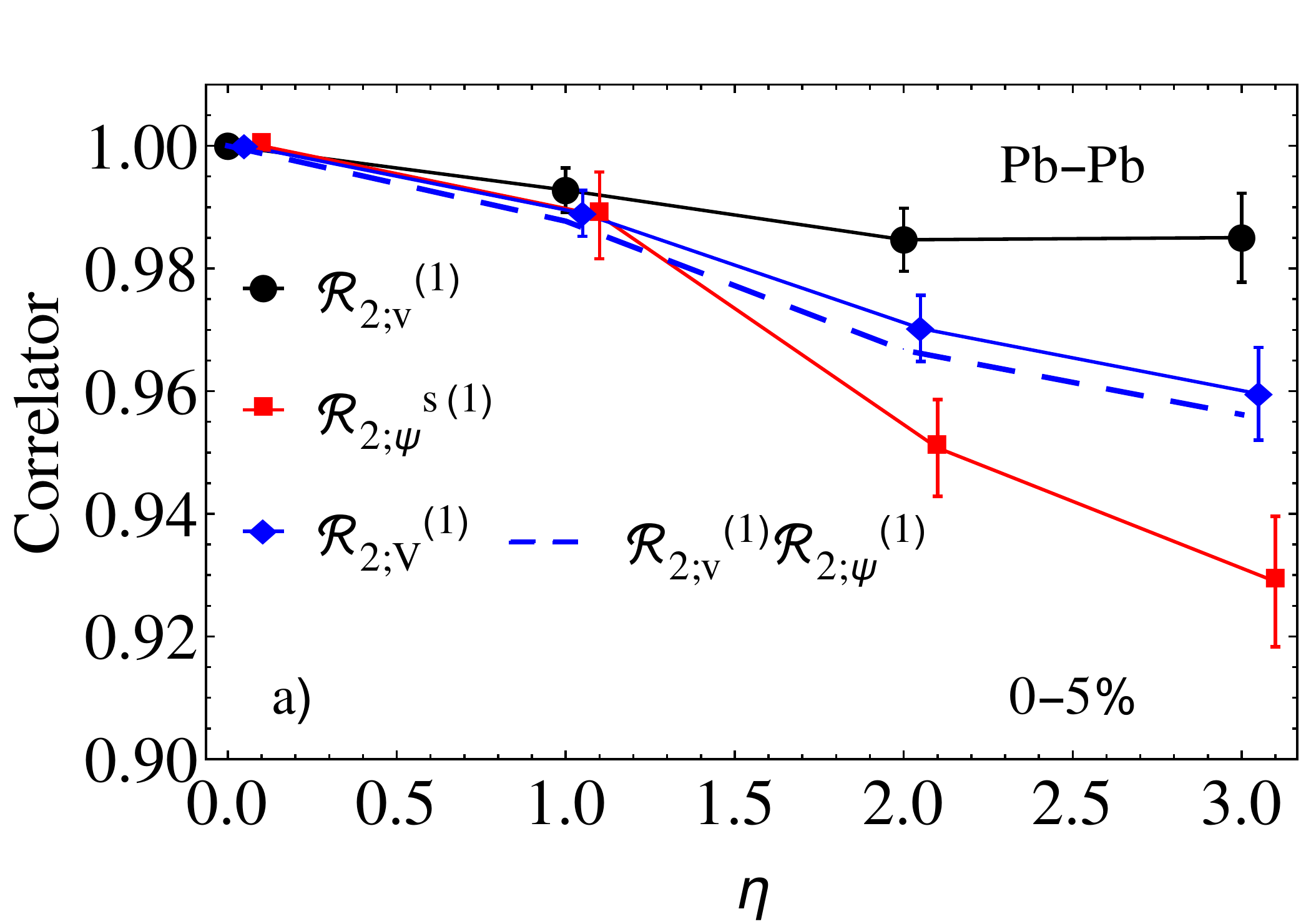} \\
	\includegraphics[scale=0.4]{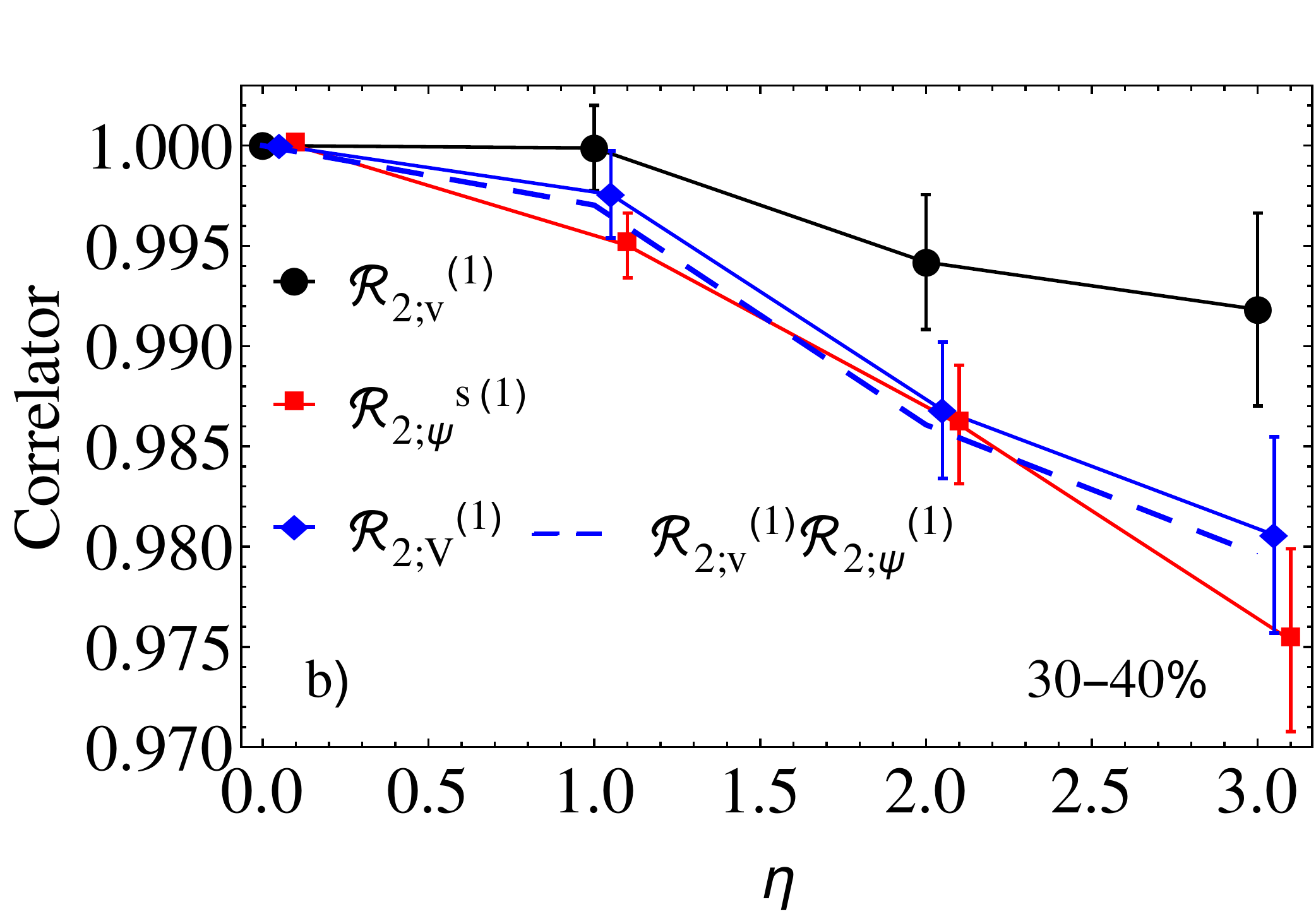}
	\caption{The factorization breaking coefficient of the elliptic flow in three pseudorapidity bins [central and semicentral collisions, panels (a) and (b) respectively]. The flow vector factorization breaking coefficients, Eq.~(\ref{eq:Rfv}), are denoted with filled circles, the flow magnitude factorization breaking coefficients, Eq.~(\ref{eq:Rfm}), are denoted with squares, and the simple flow angle factorization breaking coefficients, Eq.~(\ref{eq:fac3ang}), are denoted with diamonds. The blue dashed lines indicate an estimate of the flow vector factorization breaking coefficient, Eq.~(\ref{eq:fac3mag}), as a product of the flow magnitude factorization breaking coefficient and the  weighted flow angle factorization breaking coefficient in Eq.~(\ref{eq:Rangred}).} 
	\label{fig:3binv2}
\end{figure}
\begin{figure}[th!]
	\includegraphics[scale=0.4]{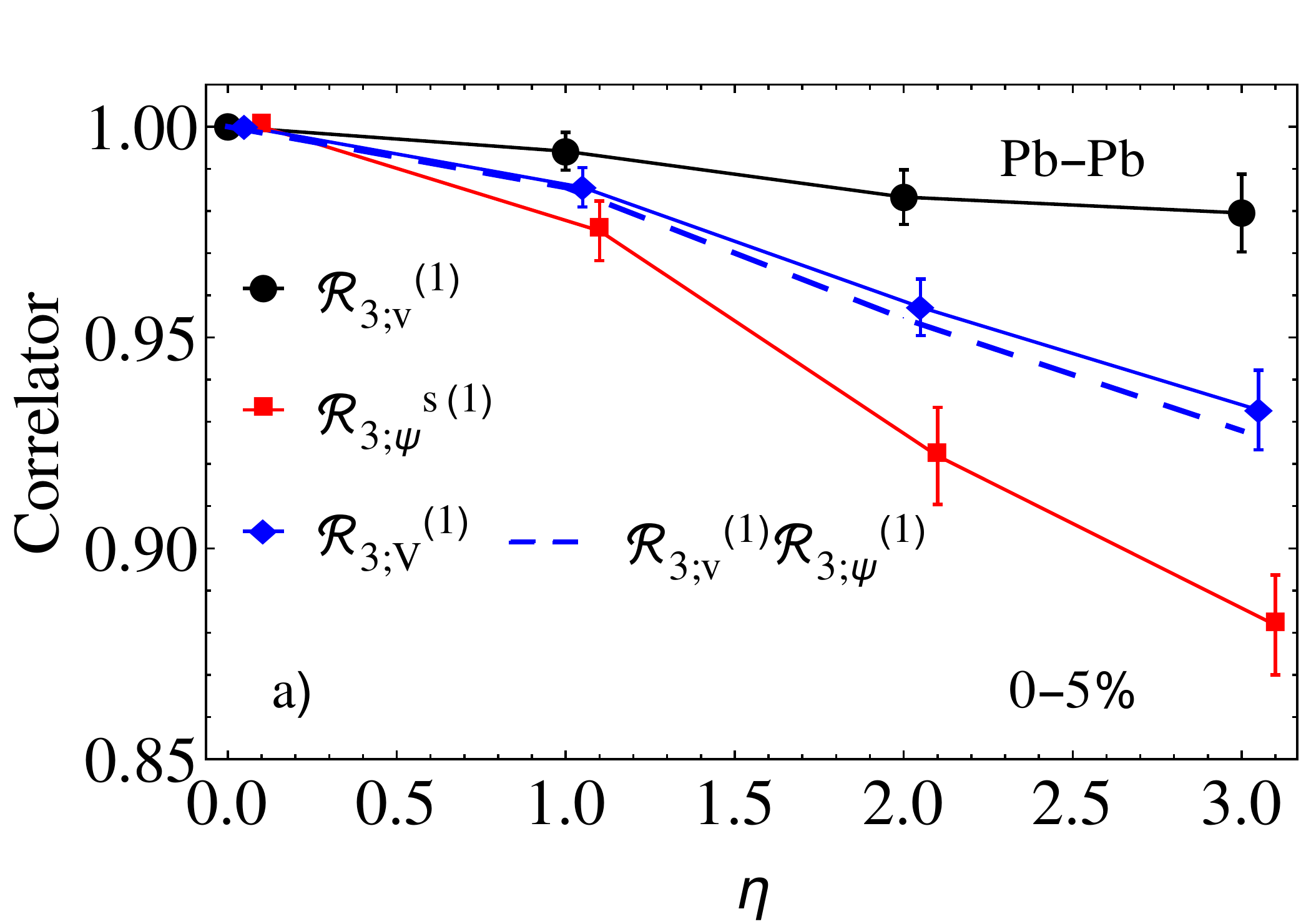} \\
	\includegraphics[scale=0.4]{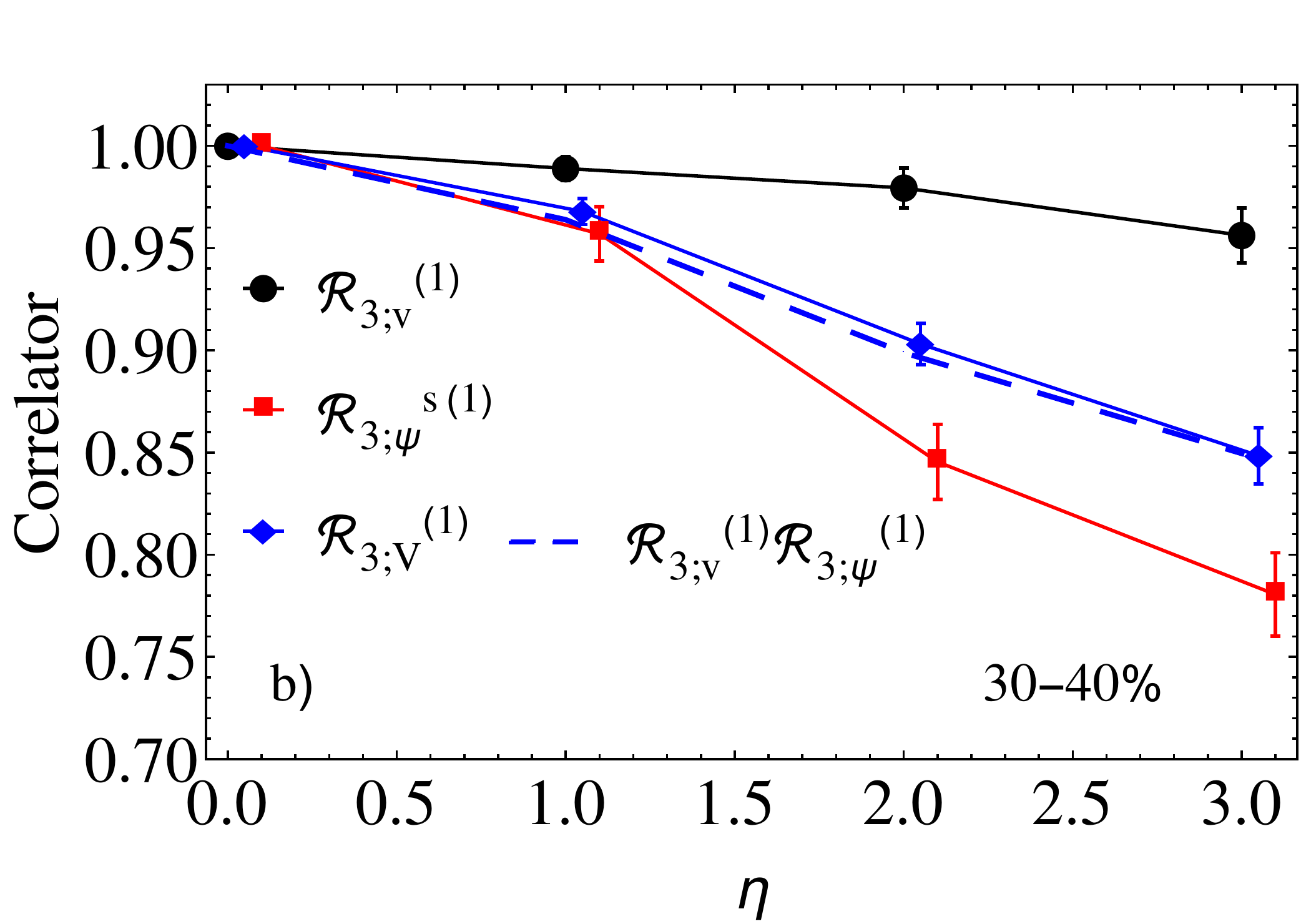}
	\caption{Same as in Fig.~\ref{fig:3binv2} but for the triangular flow.} 
	\label{fig:3binv3}
\end{figure}

The two-bin factorization breaking coefficients studied in Sec.~ \ref{sec:facbreak} are very sensitive to nonflow effects. Alternatively, a three-bin measure of flow factorization breaking in the longitudinal direction can be used  \cite{CMS:2015xmx}, Eq. (\ref{eq:r3bin}). This measure can be generalized to higher moments of the flow \cite{Jia:2017kdq,ATLAS:2017rij},
\begin{equation}
\mathcal{R}^{(m)}_{n;V}(\eta)=\frac{\langle V_n^m(-\eta)V_n^{\star m}(\eta_{ref}) \rangle}{\langle V_n^m(\eta)V_n^{\star m}(\eta_{ref}) \rangle} \ .
\label{eq:fac2flow}
\end{equation}
An analogous formula can be written for the factorization breaking coefficient of the flow magnitude as follows:
\begin{equation}
\mathcal{R}^{(m)}_{n;v}(\eta)=\frac{\langle v_n^m(-\eta)v_n^{m}(\eta_{ref}) \rangle}{\langle v_n^m(\eta)v_n^{m}(\eta_{ref}) \rangle} \ .
\label{eq:fac3mag}
\end{equation}
 A third measure can be defined to estimate the simple flow angle decorrelation:
\begin{equation}
\mathcal{R}^{s(m)}_{n;\Psi}(\eta)=\frac{\langle % v_n^{2m}
	\cos\left( m n (\Psi_n(-\eta)-\Psi_n(\eta_{ref})) \right) \rangle}
{\langle % v_n^{2m}
	\cos\left( m n (\Psi_n(\eta)-\Psi_n(\eta_{ref})) \right) \rangle}\ .
\label{eq:fac3ang}
\end{equation}

Using the local random component (K) for  the flow in a given bin $V(\eta)=V+K(\eta)$ and defining
\begin{equation} 
A_1=\frac{K(\eta)+K(\eta_{ref})}{2}\quad  \text{and} \quad\Delta_1=\frac{K(\eta)-K(\eta_{ref})}{2}\ , 
\end{equation}
and    
\begin{equation}
A_2=\frac{K(-\eta)+K(\eta_{ref})}{2}\quad \text{and} \quad\Delta_2=\frac{K(-\eta)-K(\eta_{ref})}{2}\  ,
\end{equation}
one can write:
\begin{equation}
\mathcal{R}^{(m)}_{n;V}(\eta)=\frac{\langle(V+A_2+\Delta_2)^m (V+A_2-\Delta_2)^{*m} \rangle}{\langle(V+A_1+\Delta_1)^m(V+A_1-\Delta_1)^{*m} \rangle}\ , 
\end{equation}
\begin{equation}
\mathcal{R}^{(m)}_{n;v}(\eta)=\frac{\langle|V+A_2+\Delta_2|^m|V+A_2-\Delta_2|^{m} \rangle}{\langle|V+A_1+\Delta_1|^m|V+A_1-\Delta_1|^{m} \rangle} \ .
\end{equation}
The expansion of the factorization breaking coefficients to second order in $\delta^2$, taking into account that $\langle |A|^2 \rangle = C^2 -\delta^2$, takes the form
\begin{equation}
\mathcal{R}^{(m)}_{n;V}(\eta)=1- 2 m^2 \frac{\langle v^{2m-2} \delta_2^2\rangle - \langle v^{2m-2} \delta_1^2 \rangle}{\langle v^{2m} \rangle},
\label{eq:Rfv}
\end{equation}
for the flow vector decorrelation and 
\begin{equation}
\mathcal{R}^{(m)}_{n;v}(\eta)=1-  m^2 \frac{\langle v^{2m-2} \delta_2^2\rangle - \langle v^{2m-2} \delta_1^2 \rangle}{\langle v^{2m} \rangle},
\label{eq:Rfm}
\end{equation}
for the flow magnitude decorrelation. We find that the deviation of the flow factorization breaking coefficient $R_{n;V}^{(m)}$ from $1$ is twice as large as the deviation of the flow magnitude coefficient $R_{n;v}^{(m)}$ from $1$:
\begin{equation}
[1-R_{n;v}^{(m)}(\eta)]\simeq [1-R_{n;V}^{(m)}(\eta)]/2 \ .
\end{equation}
The above relation is approximately fulfilled in numerical simulations and the experimental data \cite{Bozek:2017qir}.
Moreover, we can find the general formula for the  simple flow angle decorrelation in Eq. (\ref{eq:fac3ang}) as follows:
\begin{equation}\label{eq:46}
\mathcal{R}^{s(m)}_{n;\Psi}=1
- m^2\langle \frac{\delta_2^2}{v^2} \rangle + m^2\langle \frac{\delta_1^2}{v^2} \rangle  \ .
\end{equation}
This is not the flow angle decorrelation that is measured  in experiment. The simple flow angle decorrelation is as strong or stronger than the full flow vector decorrelation (Figs.~\ref{fig:3binv2} and \ref{fig:3binv3}).  Similarly to  the two-bin correlator, 
the correct flow angle decorrelation weighted with the power of the flow magnitude can be defined as
\begin{equation}
\mathcal{R}^{(m)}_{n;\Psi}(\eta)=\frac{\langle  v_n^{2m} 
	\cos\left( m n (\Psi_n(-\eta)-\Psi_n(\eta_{ref})) \right) \rangle}
{\langle  v_n^{2m} 
	\cos\left( m n (\Psi_n(\eta)-\Psi_n(\eta_{ref})) \right) \rangle}\ .
\label{eq:Rangred}
\end{equation}
Expanding to the order $\delta_j^2$ we find 
\begin{equation}\label{p10}
\mathcal{R}^{(m)}_{n;\Psi}\approx1-m^2\frac{\langle v^{m-2}\delta_2^2\rangle}{\langle v^{2m}\rangle}+m^2\frac{\langle v^{2m-2}\delta_1^2\rangle}{\langle v^{2m}\rangle} \ .
\end{equation}
Comparing the above results with the results for the flow vector decorrelation, Eq.~(\ref{eq:Rfv}), and flow magnitude decorrelation, Eq.~(\ref{eq:Rfm}), one observes 
\begin{equation}\label{p11}
\mathcal{R}^{(m)}_{V}\approx \mathcal{R}^{(m)}_{v}\mathcal{R}^{(m)}_{\Psi} \ .
\end{equation}
We expect that the above factorization of the flow vector coefficient works well for both elliptic and triangular flows and all centralities. This factorization is observed in numerical calculations \cite{Bozek:2017qir} (Figs.~\ref{fig:3binv2} and \ref{fig:3binv3}). It means that the relation $\mathcal{R}^{(m)}_{V}/ \mathcal{R}^{(m)}_{v}$ can be used as an estimate of the weighted flow angle decorrelation $\mathcal{R}^{(m)}_{n;\Psi}$, Eq. (\ref{eq:Rangred}). On the other hand, the simple angle decorrelation, Eq.~(\ref{eq:fac3ang}), is significantly larger.

\begin{figure}[th!]
	%\vskip 3mm
	\includegraphics[scale=0.4]{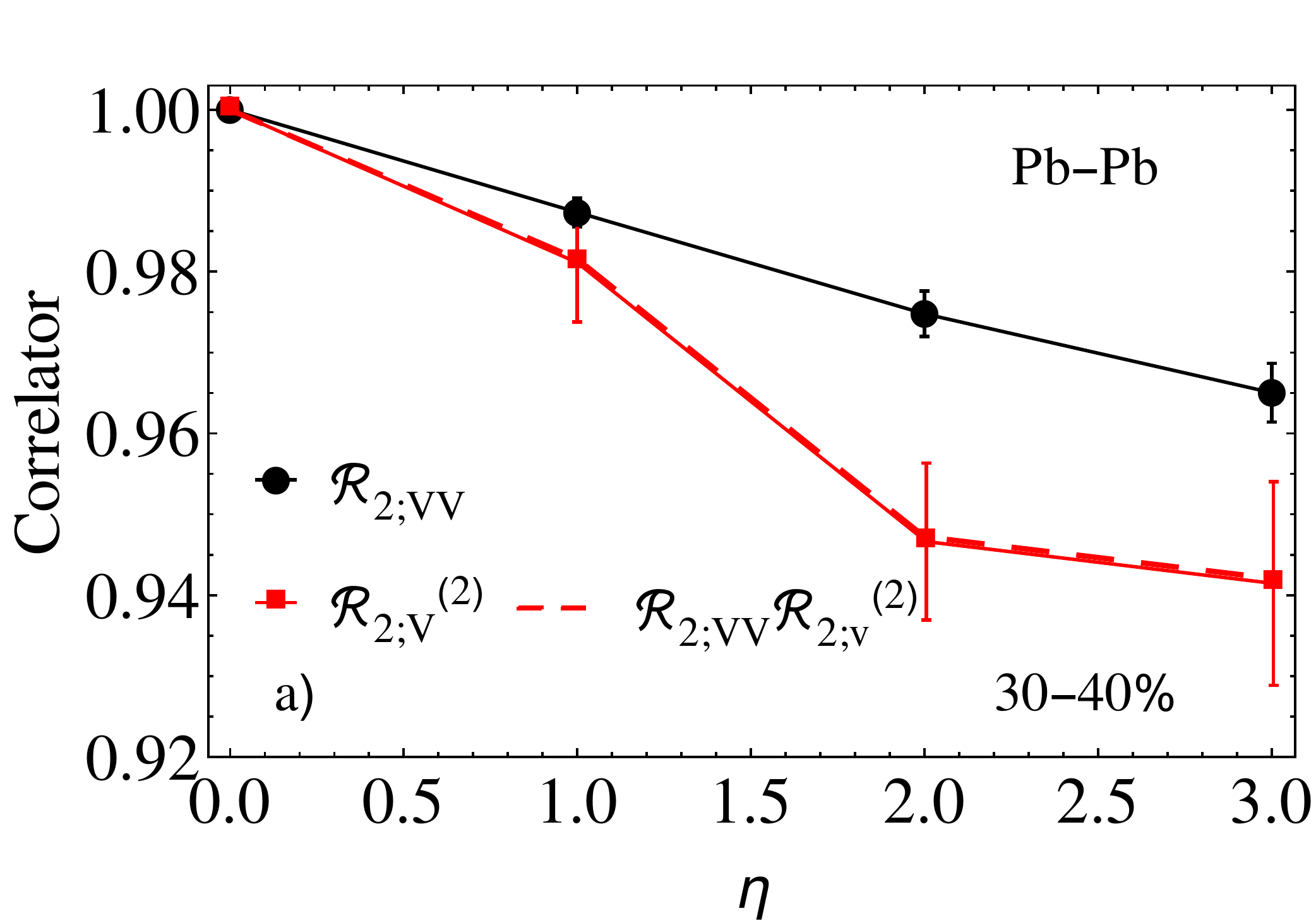} \\
	\includegraphics[scale=0.4]{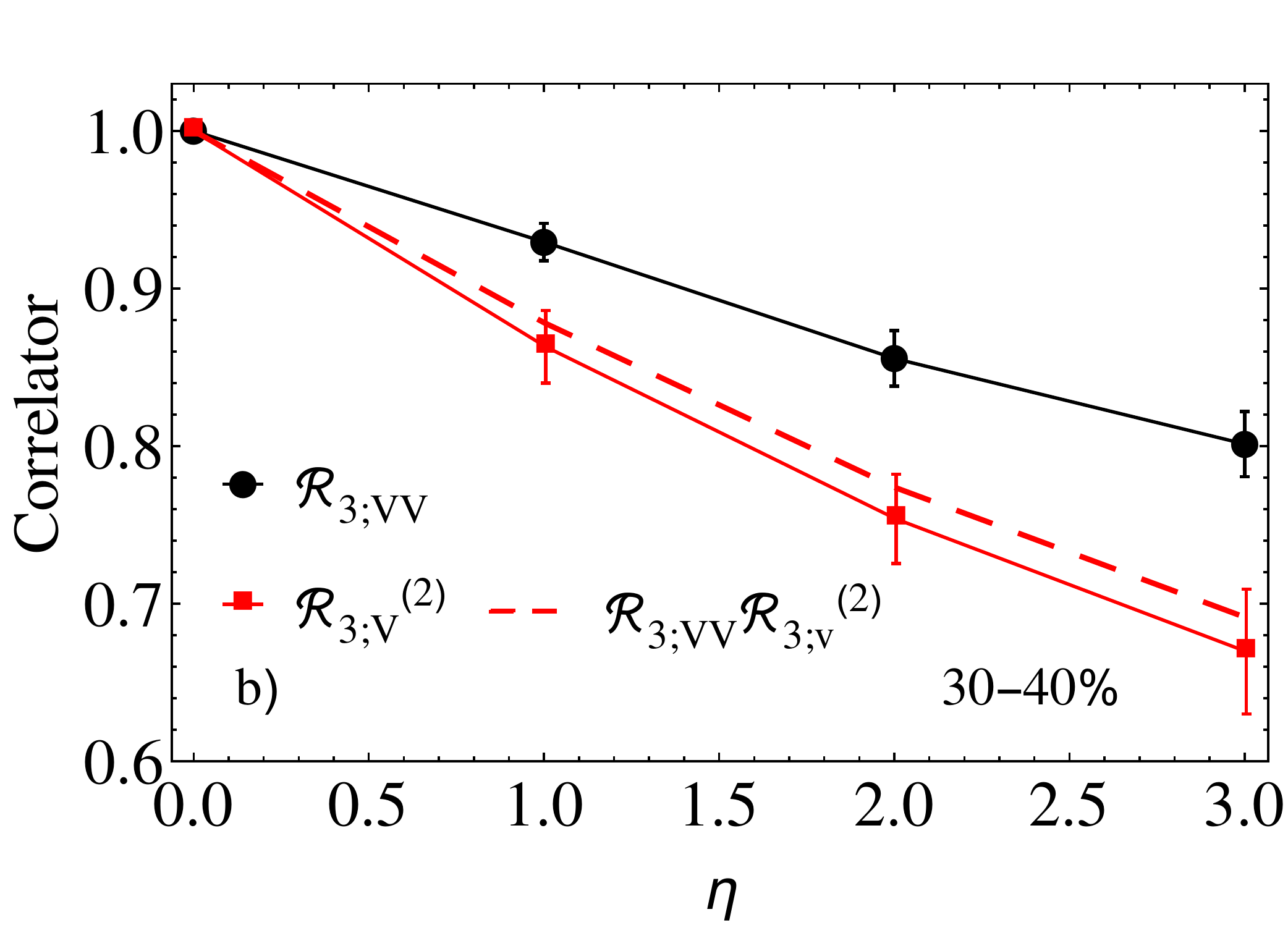}
	\caption{Comparison of three-bin (red lines) and four-bin (black lines) flow vector factorization breaking coefficients for the elliptic flow  (panel a)) and for the triangular flow (panel b)), for $30-40$\% centrality. Red dash lines show  the decomposition of  the flow vector factorization breaking coefficient   $\mathcal{R}_V^{(2)}$ into the flow magnitude and flow angle decorrelation using Eq.~(\ref{p14}).} 
	\label{fig:4binv2}
\end{figure}

%\begin{figure}[th!]
	%\vskip 3mm
%	\includegraphics[scale=0.4]{new figs/10a.pdf} \\
%	\includegraphics[scale=0.4]{new figs/10b.pdf}
%	\caption{Comparison of 3-bin (red lines) and 4-bin (black lines) flow vector factorization breaking coefficients for the elliptic flow at $0-5$\% (panel a)) and $30-40$\% (panel ()) centrality classes. Red dash lines show  the decompostion of  the flow vector factorization breaking coefficient   $\mathcal{R}_V^{(2)}$ into the flow magnitude and flow angle decorrelation using Eq.~\ref{p14}.} 
%	\label{fig:4binv2}
%\end{figure}
%\begin{figure}[th!]
%	\includegraphics[scale=0.4]{new figs/11a.pdf} \\
%	\includegraphics[scale=0.4]{new figs/11b.pdf}
%	\caption{Same as in Fig.~\ref{fig:4binv2} but for the triangular flow.} 
%	\label{fig:4binv3}
%\end{figure}

\begin{figure}[th!]
	\includegraphics[scale=0.4]{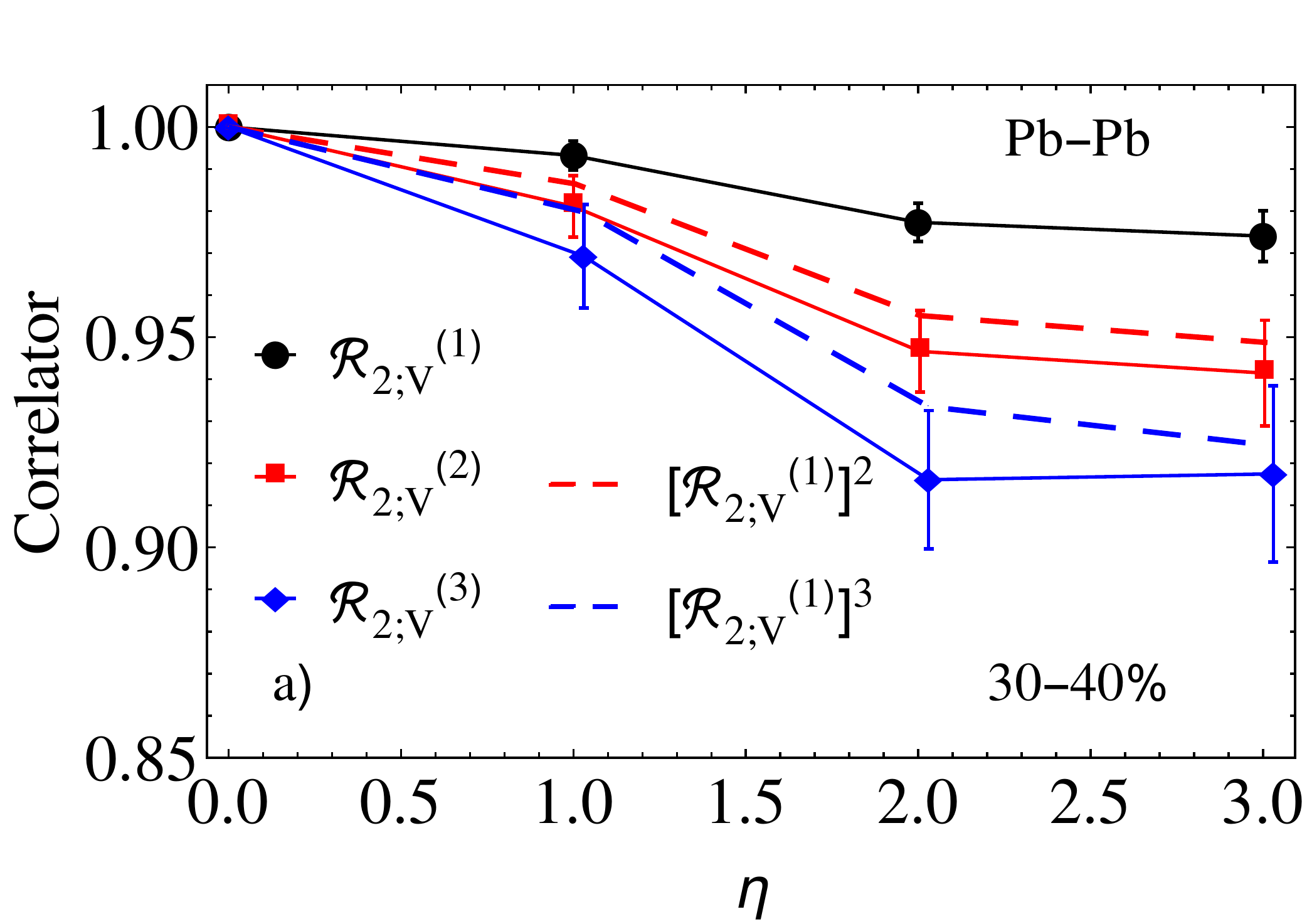} \\
	\includegraphics[scale=0.4]{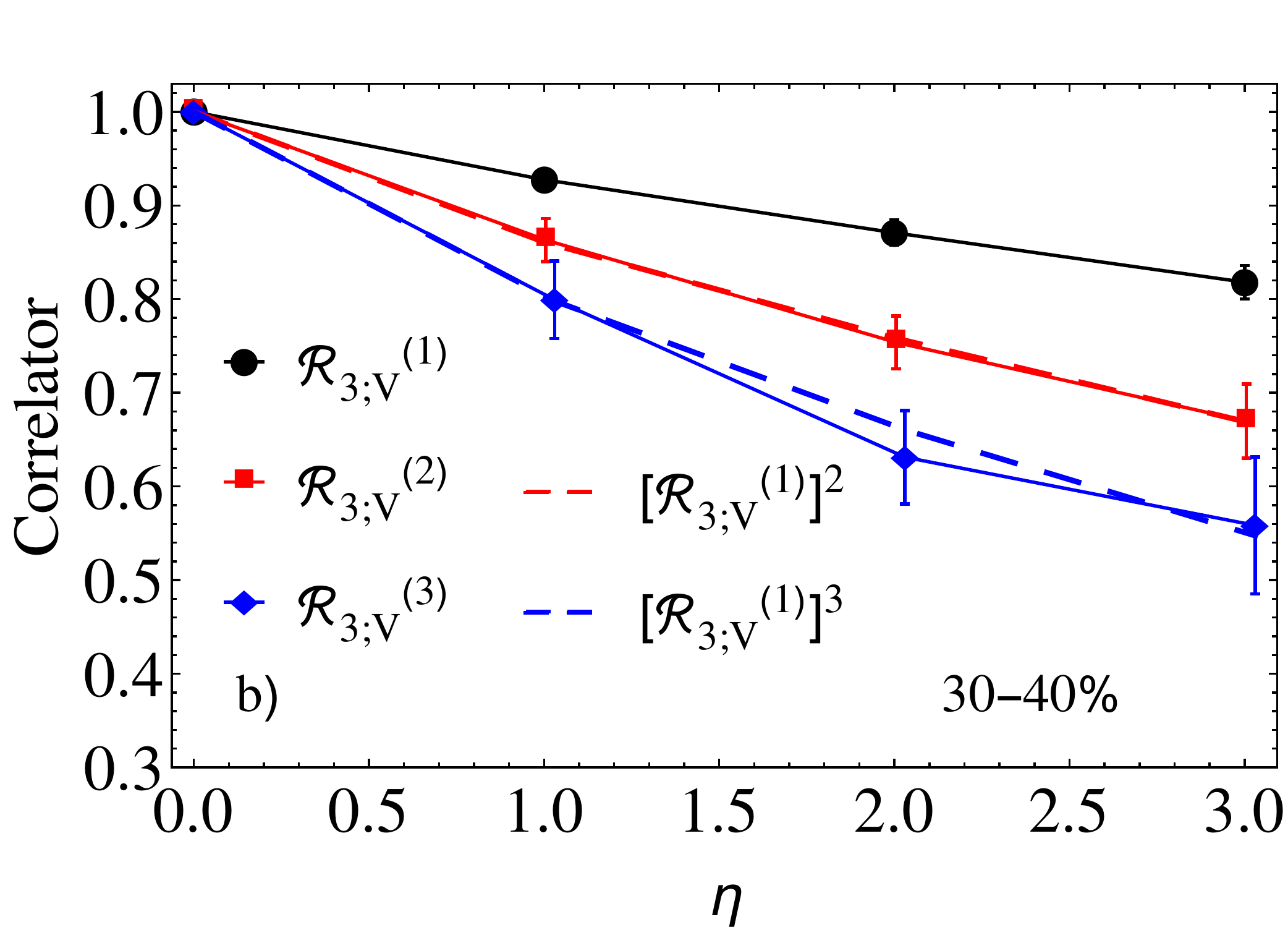}
	\caption{The factorization breaking coefficients for the first (black line and dots), second (red line and squares), and third (blue line and diamonds) powers of the flow vector $\mathcal{R}^{(m)}_{V}$ compared with the second and third powers $\Big[\mathcal{R}^{(1)}_{V}\Big]^m$  (dashed lines). Results for the elliptic flow  and the triangular flow are shown in panels (a) and (b) respectively, for the centrality range $30-40$\%.} 
	\label{fig:scale3binv2}
        \end{figure}
        
%\begin{figure}[th!]
%	\includegraphics[scale=0.4]{new figs/12a.pdf} \\
%	\includegraphics[scale=0.4]{new figs/12b.pdf}
%	\caption{Comparison of the factorization breaking coefficients for the first, second, and third moments of the elliptic flow vector $\Big[\mathcal{R}^{(1)}_{V}\Big]^m$ for $m=2-3$ (dash lines).} 
%	\label{fig:scale3binv2}
%\end{figure}
%\begin{figure}[th!]
%	\includegraphics[scale=0.4]{new figs/13a.pdf} \\
%	\includegraphics[scale=0.4]{new figs/13b.pdf}
%	\caption{Same as in Fig.~\ref{fig:scale3binv3} but for the triangular flow.} 
%	\label{fig:scale3binv3}
%\end{figure}

In the experiment, due to nonflow correlations, only the three-bin flow vector correlator $\mathcal{R}^{(m)}_{n; V}(\eta)$ can be measured. There is another practical way to estimate flow angle decorrelation at different pseudorapidities using a four-bin correlator \cite{ATLAS:2017rij},
\begin{equation}\label{p12}
\mathcal{R}_{n;VV}=\frac{\langle V_n(-\eta_{ref})V_n(-\eta)V_n^{*}(\eta)V_n^{*}(\eta_{ref}) \rangle}{\langle V_n(-\eta_{ref})V_n^*(-\eta)V_n(\eta)V_n^{*}(\eta_{ref}) \rangle}.
\end{equation}
In the random model of flow decorrelation, the above four-bin correlator takes the form  (in the lowest order in $\delta$)
\begin{equation}\label{p13}
\mathcal{R}_{n;VV}\approx1-4\frac{\langle v^{2}\delta_2^2\rangle}{\langle v^{4}\rangle}+4\frac{\langle v^{2}\delta_1^2\rangle}{\langle v^{4}\rangle} \ . 
\end{equation}
The four-bin correlator is an estimate of the weighted flow angle factorization breaking coefficient, Eq.~(\ref{eq:Rangred}). The flow vector factorization breaking coefficient $\mathcal{R}_{V}$ approximately factorizes into the flow magnitude and flow angle factorization breaking coefficient,
\begin{equation}\label{p14}
\mathcal{R}^{(2)}_{V}\approx \mathcal{R}_{n;VV}\mathcal{R}^{(2)}_{v} \ .
\end{equation}  
Numerical results from the viscous hydrodynamic model for the correlators $\mathcal{R}_{n;VV}$, $\mathcal{R}^{(2)}_{V}$, and the above factorization are shown in Fig.~\ref{fig:4binv2}. %The agreement with the experimental data is qualitatively correct.
There is a good consistency between the results of the hydrodynamic simulations and the factorization given in Eq.~(\ref{p14}) (red dashed line) in second- and third-order harmonics for both central (not shown) and semicentral collisions.

Finally, we test in the hydrodynamic model the scaling of the factorization breaking coefficients for different moments of the harmonic flow vectors:
\begin{equation}
\mathcal{R}^{(m)}_{V}\approx\left[\mathcal{R}^{(1)}_{V}\right]^m \ .
\end{equation}
Figure \ref{fig:scale3binv2}  present a comparison of the factorization breaking coefficients for the second and third moments of the flow vector $\mathcal{R}^{(m)}_{V}$ (for $m=2-3$) (red and blue lines respectively) with the respective powers of the factorization breaking coefficients of the flow vectors $\Big[\mathcal{R}^{(1)}_{V}\Big]^m$ for $m=2-3$ (red and blue dashed lines). The above relation is expected for fluctuation dominated flow (triangular flow) from Eq.~(\ref{eq:Rfv}) derived in the random model of flow decorrelation.

\section{Conclusions}

We analyze the decorrelation of the flow vectors in separate rapidity bins.
A simple random model of flow decorrelation is able to reproduce qualitatively the  scaling relations between  factorization breaking coefficients. Numerical simulations
in the hydrodynamic model show that the flow vector magnitude and flow vector angle decorrelations are approximately equal and sum up in the total flow vector decorrelation. The same relation can be obtained in a random model of flow decorrelation, where the flow in a small rapidity bin is written as a sum of the 
average flow in the event and of a random vector component. Assuming that the random component direction  is independent of the average flow and its magnitude is much smaller than the average flow, analytical expressions for the factorization breaking coefficients for flow vectors, flow magnitudes, and flow angles are given, with similar relations between them as in the full hydrodynamic simulation.

The factorization breaking coefficients for higher powers of the flow vectors shows a large deviation from $1$, reflecting the stronger decorrelation for the second or third power of the flow vectors than for the flow vectors only. In the random model of flow decorrelation this property comes from the general analytical expressions for the correlations of different moments of flow vectors. In the case when the flow is fluctuation dominated (triangular flow or elliptic flow in central collisions) the factorization breaking coefficients of different powers $m$  of the flow are related
\begin{equation}
  \rho^{(m)}(-\eta,\eta) \simeq \left[ \rho^{(1)}(-\eta,\eta) \right]^m .
  \end{equation}
The above relation is found analytically in the random decorrelation
model, as well as in  numerical simulations in the hydrodynamic model.

The flow angle decorrelation is larger if the overall flow is small.
This can be observed on an event-by-event  basis in the hydrodynamic simulations and is encoded
in the analytical expressions obtained in the random model of  flow decorrelation.
As a consequence the flow decorrelation is  larger for the triangular flow and for the elliptic flow in central collisions than for the elliptic flow in semicentral collisions.

Analytical expression of the  factorization breaking coefficients are given for the three- and four-bin measures of flow decorrelation in pseudorapidity. Such measures are used in experimental analyses in order to reduce nonflow effects.
The random model of flow decorrelation reproduces qualitatively the relation observed in experimental data and in hydrodynamic simulations:
	
\begin{itemize}
\item{} the flow angle decorrelation is approximately one-half of the  flow vector decorrelation,
  \item{} the decorrelation of the second or third powers of the flow is given as the second or third powers of the flow vector decorrelation, when the overall flow is fluctuation dominated.
\end{itemize}

The proposed random model of flow decorrelation in pseudorapidity is surprisingly simple, yet it reproduces qualitatively  a number of phenomenological relations observed in experimental data or in realistic hydrodynamic simulations. The model can serve as a way to understand the fluctuations inherent in models of heavy-ion collisions.
Moreover, any deviations from these simple scalings in the experimental data could suggest the
existence 
of specific correlations between the average flow and the random flow
decorrelations, which could be interesting in
the study
of realistic initial conditions for the hydrodynamic simulations of heavy-ion
collisions. The possible existence of a similar simple description of  the decorrelation of the flow  in different transverse momentum bins is open and left for future studies.

\section*{Acknowledgments}
P.B. acknowledges support from the National Science Centre  Grant  2018/29/B/ST2/00244.
H.M.~thanks CERN-TH group for  support. H.M.~is funded by the Cluster of Excellence {\em Precision Physics, Fundamental Interactions, and Structure of Matter} (PRISMA$^+$ EXC 2118/1) funded by the German Research Foundation (DFG) within the German Excellence Strategy (Project ID 39083149).
	
\bibliography{hydr.bib}

\end{document}